\documentclass[12pt]{article}
\usepackage{amsmath,amsfonts,amssymb,color,a4,epsfig,graphics}
\usepackage[latin1]{inputenc}

\newcommand{\R}{{\mathbb{R}}}
\newcommand{\C}{{\mathbb{C}}}
\newcommand{\T}{{\mathbb{T}}}
\newcommand{\Z}{{\mathbb{Z}}}
\newcommand{\N}{{\mathbb{N}}}

\def\Di {\displaystyle}

\def\ha{\frac{1}{2}}

\def\pa{\partial}
\def\ra{\rightarrow}
\def\preuve{{\it Proof.-- ~}}
\def\ga{\alpha}

\def\gd{\delta}
\def\ge{\varepsilon}

\def\gg{\gamma}

\def\gl{\lambda}
\def\go{\omega}

\def\gs{\sigma}
\def\gt{\tau}

\def\gO{\Omega}

\def\gF{\Phi}
\def\gG{\Gamma }

\def\gS{\Sigma}
\def\gL{\Lambda}

\def\finpreuve{\hfill$\square $}
\newtheorem{defi}{Definition}[section]

\newtheorem{lemm}{Lemma}[section]
\newtheorem{prop}{Proposition}[section]
\newtheorem{rem}{Remark}[section]
\newtheorem{coro}{Corollary}[section]
\newtheorem{theo}{Theorem}[section]
\newtheorem{exem}{Example}[section]

\newenvironment{demo}{\noindent {\it Proof.--}
      \begin{quotation}\noindent}{\end{quotation}\hfill$\square $}

\renewcommand{\footnote}[1]{\footnotetext{#1}}

\begin{document}

\footnote{ \textbf{ Math Subject Classification (2000):}
 05C63, 05C50, 05C12, 35J10, 47B25.}
\title{Scattering theory  for graphs isomorphic to a 
homogeneous tree at infinity}
\author{Yves Colin de Verdi\`ere\footnote{Grenoble University,
Institut Fourier,
 Unit{\'e} mixte
 de recherche CNRS-UJF 5582,
 BP 74, 38402-Saint Martin d'H\`eres Cedex (France);
{\tt yves.colin-de-verdiere@ujf-grenoble.fr};
{\tt http://www-fourier.ujf-grenoble.fr/$\sim $ycolver/}
}\\
 Fran{\c c}oise Truc
\footnote{Grenoble University, Institut Fourier,
Unit{\'e} mixte
 de recherche CNRS-UJF 5582,
 BP 74, 38402-Saint Martin d'H\`eres Cedex (France);
{\tt francoise.truc@ujf-grenoble.fr};
{\tt http://www-fourier.ujf-grenoble.fr/$\sim$trucfr/}
}}
%\date{}
\maketitle
%%%%%%%%%%%%%%%%%%%%%%%%%%%%%%%%%%%%%%%%%%%%%%%%%%%%%%%%%%%%%%%%%%%%%%%%%%%%%%%%%%%%%%%%%%%%%%%%%%%%%%%%%%%%%%%%%%%%%%%%%%%%%%%%%%%%%%%%%%%%%
\begin{abstract}
We describe the spectral theory of the adjacency operator
of  a graph which is isomorphic to a regular tree at infinity.
Using some combinatorics, we reduce the problem to a scattering
problem for a finite rank perturbation of the adjacency operator
on a regular tree. We develop this scattering theory using
the classical recipes for Schr\"odinger operators in Euclidian spaces.
\end{abstract}

\section{Introduction}
The aim of this paper is to describe in an explicit way the spectral
theory
of the adjacency operator on an infinite graph $\gG$  which, outside of a
finite  sub-graph  $\gG_0$,  looks like a regular tree $\T_q$
 of degree $q+1$.
We mainly adapt   the case of the Schr\"odinger 
operators as presented in \cite{RS,Ike}.
The proofs are often simpler here  and the main results
are similar. This paper can be read
as an introduction to the scattering
theory for differential operators on smooth manifolds.
Even if we do  not find our  results in the literature, there is probably 
nothing really new for experts in the scattering theory of  Schr\"odinger 
operators, except  the combinatorial part in Section \ref{sec:combi}.  

The main result is an explicit spectral decomposition: the Hilbert
space $l^2(\gG)$  splits into a sum of two invariant subspaces 
$l^2(\gG)={\cal H}_{\rm ac} \oplus {\cal H}_{\rm pp}$.
The first one is an absolutely continuous part isomorphic to
 a closed sub-space of
that of the regular tree of degree $q+1$,  while 
the second one is finite dimensional and we have an upper bound
 on its dimension.
The absolutely continuous part of the spectral decomposition
is given in terms of explicit generalized
eigenfunctions whose
behavior at infinity is described  in terms of a scattering matrix.
  
We first introduce the setup, then we recall the spectral decomposition  of 
the adjacency operator $A_0$ of a regular tree $\T_q$ by using the
Fourier-Helgason
 transform.
In Section 3,  we consider a Schr\"odinger 
operator $A = A_0 + W$ on $\T_q$, where $W$ is a  compactly supported
non
 local potential. We build the generalized eigenfunctions for $A$, 
define a  deformed Fourier-Helgason transform and 
get a spectral decomposition of $A$ (Theorem \ref{fhtil}). In section 4,
 we derive a similar 
 spectral decomposition of the adjacency operator of
any graph $\gG$  asymptotic to a regular tree $\T_q$ by proving
 the following combinatorial result (Theorem 4.2):
any such graph $\gG$ is isomorphic to
a connected component of a graph ${\hat\gG}$
which is obtained from $\T_q$ by a finite number of
modifications. This implies  that the adjacency
 operator of 
 ${\hat\gG}$ is a finite rank perturbation of the 
adjacency
 operator of  $\T_q$.
In section 5, we investigate some consequences of 
 the scattering theory developed in section 3: 
 we write
the point-to-point correlations of scattered waves in terms of the
 Green's function,
 we define the transmission
 coefficients,
connect them to the scattering matrix, and get an explicit expression
 of them in terms of
a Dirichlet-to-Neumann operator.  For the sake of clarity, this part
has been
 postponed,
since it is not necessary to prove Theorem 4.2.
 
\section{The setup: graphs  asymptotic to a
 regular  tree  }

Let us consider a  connected graph $\gG =(V_\gG,E_\gG)$ with $V_\gG$
the set of {\it vertices} and $E_\gG$ the set of {\it edges}. We write
 $ x\sim y$ for  $\{x,y\}\in E_\gG $.

\begin{defi} \label{astl}
Let $q\geq 2$ be a fixed integer. We say that the infinite
connected  graph $\gG$ is
  {\rm  asymptotic to a regular tree of degree $q+1$}
 if there exists a finite sub-graph $\gG_0$ of $\gG$  such that 
$\gG':=\gG\setminus  \gG_0$ is a disjoint union of a finite number
of trees $T_l,~l=1,\cdots, L,$ rooted at a vertex $x_l$ linked
 to $\gG_0$ and so that
all vertices of $T_l$ different from  $x_l$ are of degree $q+1$. 
The trees $T_l,~l=1,\cdots, L,$ are called the {\rm ends} of $\gG$. 

Equivalently, $\gG $ is infinite,  has a finite number of cycles
 and a maximal
sub-tree of $\gG $ has all vertices of degree $q+1$ except a finite
number of them.
\end{defi}
\begin{figure}[hbtp]
\leavevmode \center
\input{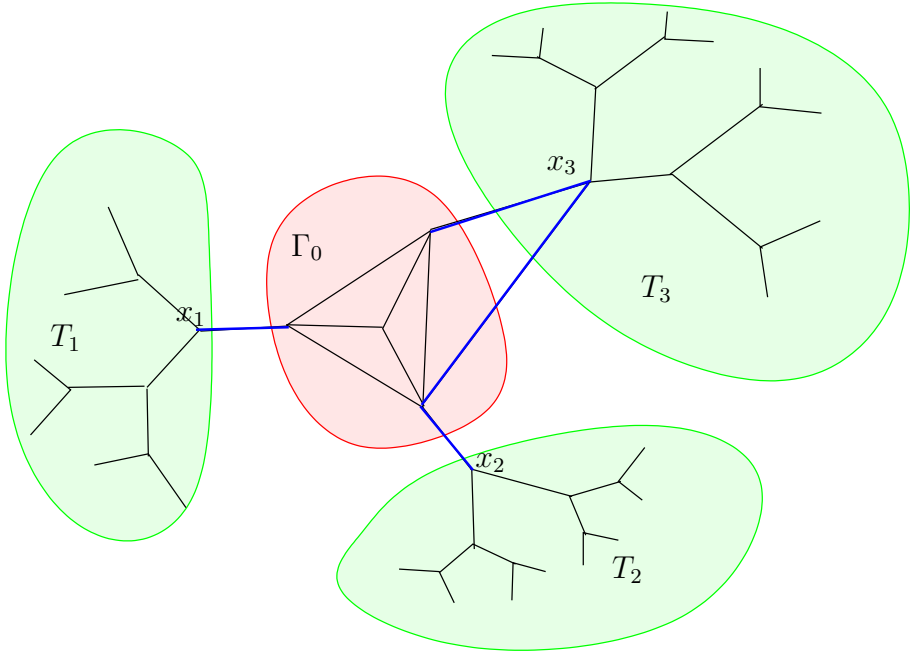}
\caption{{\it \small  A graph $\Gamma $ asymptotic to a regular 2-tree
with $L=3$; the edge boundary $\pa_e \Gamma_0$ has  4 edges.}}
 \label{fig:Gamma}
\end{figure}

\begin{defi} \label{bdry}
We define the  {\rm edge boundary } $( \pa_e\gG_0)$ of $\gG_0$ as the set
 of edges of  $\gG $ connecting
a vertex of $\gG_0$ to
a vertex of  ${\gG'}$, namely one of the $x_l$'s.
We denote by  $|x|_{\gG_0}$ the combinatorial distance of $x\in V_{\gG} $
to $\gG_0$.
\end{defi}
In particular, for $l=1,\cdots, L$, $|x_l|_{\Gamma _0}=1$.

The space of complex-valued functions on  $V_\gG$ is denoted
$$C(\gG)=\left\lbrace f:V_\gG \longrightarrow \C \right\rbrace $$
and $C_0(\gG)\subset C(\gG) $ is the subspace of functions with finite support.
We define also
  $$l^2(\gG )=\lbrace f\in C(\gG );\; \underset{x\in
V_\gG }{\sum} |f|^2(x) <\infty \rbrace.$$
It is a Hilbert
space when equipped with  the inner product:
\[ \langle f,g \rangle =\sum_{x\in V_\gG }
 \overline {f(x)}. g\left( x\right)~.\]
Let us emphasize that we take the physicist's notation,
 as in \cite{RS} for example: our 
 inner product is  conjugate-linear in the first vector and
linear
 in the second.
\newline
On $C_0(\gG )$, we define  the adjacency operator $A_\gG$
%\cite{If $B$ is a linear operator from $C_0(\gG)$into $C(\gG)$,we denote by $[B](x,y),~x,y\in V_\gG $ its matrix } 
by the formula: 
 \begin{equation} \label{adja}
 \left( A_\gG f\right)\left( x\right)
 = \sum_{y\sim x}f\left(y \right)
 \end{equation}
The operator  $A_\gG$  is bounded  on $l^2(\gG )$
if and only if the degree of the vertices of $\gG $ is bounded,
 which is the case here.
In that case, the operator $A_\gG$ is self-adjoint; otherwise,
the operator $A_\gG$ defined on $C_0(\gG)$ 
could have several self-adjoint extensions.

For any $\gl $ outside the 
spectrum of $A_\gG $, we denote by $R_\gG (\lambda ):l^2(\Gamma )
\ra l^2 (\Gamma )$ 
the resolvent $(\gl -A_\gG)^{-1}$
and by $G_\gG (\gl,x,y)$ with $x,y \in V_\Gamma $
 the matrix of $R_\gG (\lambda )$,
also called the {\it Green's function}.

\section{The spectral decomposition of the
adjacency matrix  of the   tree $\T_q$ and the Fourier-Helgason
transform}

\subsection{Points at infinity}

Let  $\T _q=(V_q,E_q) $ be the regular tree of  degree $q+1$
and let us choose   an origin, also called a root,
  $O$. We denote by $|x|$ the combinatorial distance of the vertex
$x$ to the root. 
The set of points at infinity denoted  $\Omega_O $ is the set of
infinite simple paths starting from $O$.
We will say that a sequence $y_n \in V_q$ tends to $\omega
\in\Omega_O $ if, for $n$ large enough, $y_n$ belongs to the
path $\omega $ and is going to infinity along that path.
If $x$ is another vertex 
of $V_q$, the sets $\Omega _O$ and $\Omega _x$ are canonically identified
by considering paths which co\"incide far from $O$ and $x$.
There is a canonical probability measure $d\sigma _O $  on $\Omega _O $:
$d\sigma_O $ is the unique probability measure on $\Omega _O$
which is 
invariant by the automorphisms of $\T_q$ leaving $O$ fixed. 
Later on we will always denote by $\Omega $ the set of points at infinity,
because the root is fixed. 
\begin{figure}[hbtp]
\leavevmode \center
\input{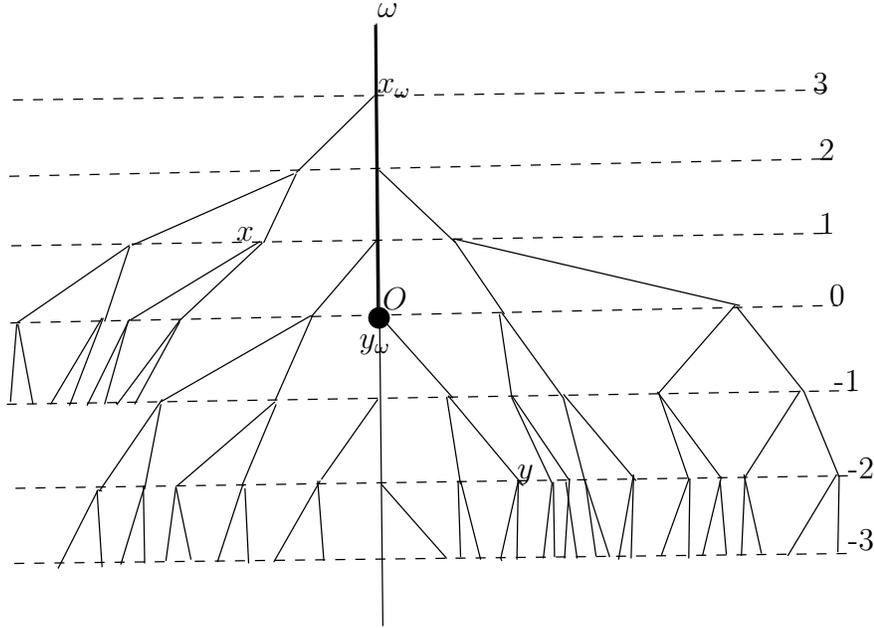}
\caption{{\it \small  A regular tree with $q=2$ and some level-sets of a 
Buseman function}}
 \label{fig:horo}
\end{figure}
For the tree $\T_q$, the {\it Busemann function} $x\ra b_\go (x)$
 associated to the point 
 $\go \in \Omega _O$ is defined as follows:
let us denote by $x_{\go}$
  the last point lying on  $\go$ in the geodesic
path joining $O$ to $x$, (take $x_\go =O$
in the case where $O$ belongs to the geodesic from $x$ to $\go$),
 and let us set 
 $b_{\go}(x)= |x_{\go}|-
d(x,x_{\go})$. 
 The level sets of $b_\go $ are the {\it horocycles}
associated to $\go $. We notice  that 
 the function $b_\go (x)$ increases by one for one of $x$'s
 neighbors, namely
the one of the ray from $x$ to $\go$,
and decreases for  the others. Thus the function $b_\go (x)$ goes to $+\infty $
as $x$ tends to $\go $. 
As $x$ tends to $\go' \ne \go $, the function  $b_\go (x)$ tends
 to $-\infty $, whereas the quantity $ b_\go (x)+|x| $
 remains bounded, since it
 tends to $ 2 |x_{\go}|$.
%dessin ...

\subsection{The spectral Riemann surface}

Let us define the Riemann surface $S=\R/\tau \Z \times i\R $
with $\tau = 2\pi/ \log q $.
We denote by $S^0 =\R /\tau \Z $ the circle
$\Im s=0$, and we set 
$I_q:=[-2 \sqrt{q},+2 \sqrt{q}]$.
\begin{defi}
For  any $s \in S$,  we set : $\gl_s=q^{\ha +is}+q^{\ha -is} \ .$
\end{defi}
\begin{prop} The
 map ${\tilde\gL}:$
$s\ra \gl_s$ is holomorphic from $S$ to $\C $. 
It maps bijectively the physical sheet $S^+ =\{ 
s\in S~|~ \Im s >0 \}$  onto
$\C \setminus I_q $.
 By this map  
the circle $S^0$ is a double covering of $I_q$. 
\end{prop}
\begin{defi}
If $J$ is a subset of $\C $, we denote by 
$\hat{J}$ the {\it pre-image} of $J$ by the map ${\tilde\gL}$,
 i.e. $\hat{J}$ is the subset of $S$ defined
by  $\hat{J}:=\{ s \in S~|~\gl_s \in J \}~$.
\end{defi}
% where, as we will see, $I_q$ is the spectrum of $A_0:=A_{\T_q}$, namely the 
%real segment $[-2 \sqrt{q},+2 \sqrt{q}]$.
\begin{figure}[hbtp]
\leavevmode \center
\input{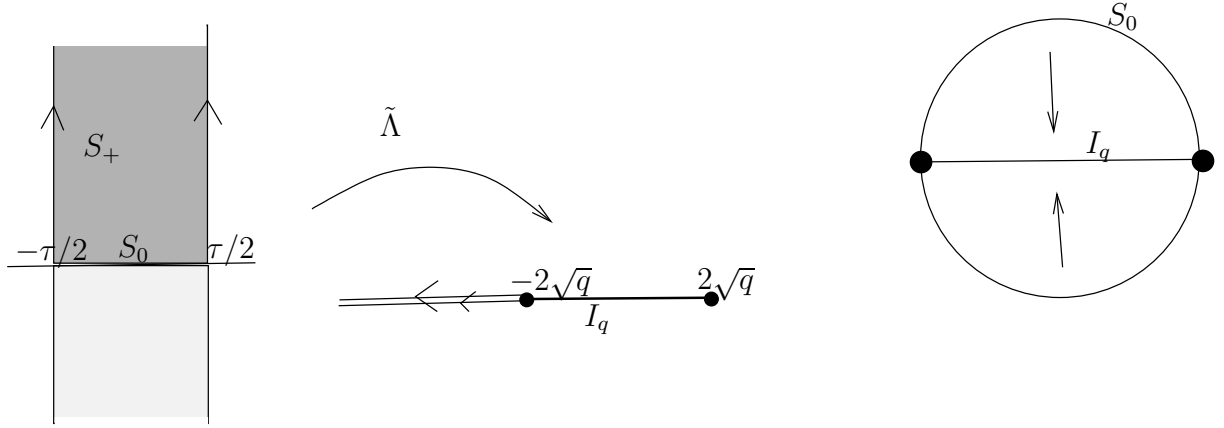}
\caption{{\it \small The surface $S$, the map $\tilde{\Lambda}$ from
    $S$
to $\C$, and the double cover of $S_0$ over $I_q$.}}
 \label{fig:S}
\end{figure}

\subsection{Calculation of the Green's function}
The results of this section are classical, see for example
the paper by P. Cartier \cite{Cartier}. 
We denote by $A_0$ (resp. $G_0$) the adjacency operator (resp. the
Green's function) on 
$\T_q$.
We will compute explicitly
 ${ G_0}(\gl,x,y)$.
Let us recall that the regular tree is 2-point regular: for any
$x,y,x',y'\in V(\T_q)$ so that 
$d(x,y)=d(x',y')$, there exists an automorphism
$J$ of $\T_q$ so that $J(x)=x'$ and $J(y)=y'$.
The Green's function $G(\gl,x,y)$ satisfies
$G(\gl ,Jx,Jy)=G(\gl,x,y)$ for any automorphism $J$ of $\T_q$.
Hence, $G(\gl,x,y)$ is a function of the distance $d(x,y)$.
It is  therefore enough to compute
  $ { G_0}(\gl,O,x)$
 for an   $x \in V_q$, that is the value $f(x)$ of the
$l^2$ 
solution of
\begin{equation} \label{resp}
 (\lambda-A_0)f= \delta_O \ ,
\end{equation} 
where  $f(x)$ depends only on the distance
 $|x|$ to the origin $O$.  So we
set $f(x)=u_k$ if $|x| =k$, $k\in\N$,
 and rewrite equation (\ref{resp}) as follows:
 \begin{description}
   \item i) $\lambda u_k -q u_{k+1} -u_{k-1} =0 $ \quad for\quad $k\geq 1$
   \item ii) $\lambda u_0 -(q+1) u_1  =1  $
    \item iii) $\sum_{n=0}^{\infty} (q+1)q^{n-1} u_n^2 < +\infty$
\end{description}
The last condition stands for $f$ to be  in $l^2(\T_q)$. 
\begin{itemize}
\item 
If $\gl\notin I_q$, the equation 
\begin{equation*}
q\alpha^2-\lambda \alpha +1=0
\end{equation*}
 admits an unique solution $\ga $  such that $|\alpha | <1/\sqrt q$.
From i) and iii), we get that $ u_k= C\alpha^k$
and the constant $C$ is determined  by ii) :
\begin{equation*}
C=C_{\gl}=\frac{1}{\lambda-(q+1)\alpha}\ .
\end{equation*} 
Therefore we have
\begin{equation*} \label{msp}
  G_0(\gl,O,x)=\frac{2q \alpha^{|x|}}{\lambda(q-1)+(q+1)F(\lambda)}
\end{equation*}
where $F(\gl)$  denotes 
the determination
 of $\sqrt{\gl^2-4q}$ in $\C \setminus  I_q$  equivalent to $\gl$ as
 $\gl$ tends to infinity. Thus using the invariance of the Green's function 
by the group of automorphisms of the tree, we see that the Green's
function $G_0(\lambda ,x,y)$ 
is a function of the distance $d(x,y)$ and 
 we have, for any $x,y \in V(\T _q)$,
\begin{equation} 
 G_0(\gl,x,y)=  C_{\gl} \alpha^{d(x,y)}~. 
\end{equation}
The operator of matrix $G_0(\gl,.,.)$ is clearly bounded
in $l^2(\T_q)$ and $\gl $ is not in the spectrum of $A_0$.
\item If $\gl \in I_q$,  there is no $l^2$ solution of Equation
(\ref{resp}). Therefore we cannot solve 
$(\lambda -A_0)f=\gd_O$ in $l^2$,   the resolvent does not 
exist  and $\gl $ is in  the spectrum of $A_0$.
\end{itemize}

Using the parameter $s \in S^+ $,
we have \begin{equation*} \label{mspa}
\alpha =q^{-\ha +is},\quad \quad C_{\gl_s}:=C(s) =
\frac{1}{q^{\ha -is}-q^{-\ha +is}}\quad
{\rm and} \quad F(\gl_s)=q^{\ha -is}-q^{\ha +is}\ .
\end{equation*}
\begin{theo}\label{greenarbre}
The spectrum of $A_0$ is the interval $I_q=[-2\sqrt{q},+2\sqrt{q}]$.

The Green's function of the tree $\T_q$ is given, for $s\in S^+$  by
\begin{equation}\label{mspi}
 G_0(\gl_s,x,y)=C(s)q^{(-\ha+is) d(x,y)}=
\frac{q^{(-\ha+is) d(x,y)}}{q^{\ha -is}-q^{-\ha +is}}~.
\end{equation}
As a function of $s$, the Green's function extends meromorphically
to $S$ with two  poles $-i/2 $ and $-i/2+ \tau/2  $. 

Moreover we have, for any $x \in V_q$ and any $y$ belonging to the
ray from $x_\go $ to $\go$,  
\begin{equation} \label{futtt}
 G_0(\gl_s,x,y) = G_{rad}(\gl_s, y) q^{(\ha -is)b_{\go}(x)}
 \end{equation}
with \begin{equation} \label{rad}
  G_{rad}(\gl_s, y) = C(s) q^{(-\ha +is)|y|} 
 \end{equation}
\end{theo}

\begin{demo}
 The last result comes from the
definition  $b_{\go}(x) = |x_{\go}|-
d(x,x_{\go})$.
\end{demo}

\subsection{The density of states}\label{ss:dos}
Let us recall how to introduce a notion of spectral measure
 (also called density of states)
on the graph $\gG$. For a given continuous
 function $\phi:\R \rightarrow \R$, 
we associate by the functional calculus an operator
 $\phi(A_\gG )$ on  $l^2(\gG)$, which has 
 a matrix $[\phi(A_\gG )](x,x')$. We consider then,
 for any $x\in V_\gG$,  the linear form on $C(\R,\R)$
$$ L_{x}(\phi)= [\phi(A_\gG)](x,x) \ .$$
$L_{x}$ is positive and satisfies  $L_x(1)=1$,
 so we have $ L_{x}(\phi)=\int _\R \phi de_x$
 where  $de_x$ is a probability measure on $\R$,
supported by the spectrum of $A_\gG$  which is called 
the spectral measure of $\gG$ at the vertex $x$.

The density of states of $\T_q$  is given by the
\begin{theo}\label{cdv}(See for example \cite{CdV2})
The spectral measure $de_x$ of  $\T_q$ 
 is independent of the vertex $x$ and is given by
\begin{equation} \label{msa}
 de_x (\gl):= de (\gl)=\frac{(q+1)\sqrt{4q-\gl^2}}{2\pi
 \left((q+1)^2-\gl^2\right)} d\gl
 \end{equation}
 \end{theo}

\begin{demo} For the sake of clarity, we recall the main ingredients: 

1) an explicit computation of the diagonal entries  of the 
Green's function
\begin{equation*} \label{ms}
 G_0(\gl,x,x)=\frac{2q}{\lambda(q-1)-(q+1)F(\lambda)}
\end{equation*}
where $F(\gl)$ denotes as previously 
the determination of
 $\sqrt{\gl^2-4q}$ in $\C \slash I_q$ (with $I_q =[ -2\sqrt q,2\sqrt
 q] $)
 equivalent to $\gl$ 
for great values of $\gl$.

 2) The expression of the spectral measure via Stone formula
\begin{equation} \label{sff}
 de (\gl)=\frac{-1}{2i\pi}\left(G(\gl+i0,x,x)
-G(\gl-i0,x,x)\right)dt\ .
\end{equation}\end{demo}

 The previous density of states is the  weak limit for the densities of
a graph asymptotic to a regular tree. More precisely we have
\begin{theo}\label{dec1}
Let $\gG$ be as in definition \ref{astl}. Consider  the adjacency operator $A_\gG$
%\cite{If $B$ is a linear operator from $C_0(\gG)$into $C(\gG)$,we denote by $[B](x,y),~x,y\in V_\gG $ its matrix } 
defined by (\ref{adja}), denote $A:=A_\gG$ for simplicity. When  $x$ tends to infinity,
 the densities of states $d\mu^A_{x}(\gl)$
 of  $\gG$ converge weakly
 to the density of states $ de(\gl)$  of 
 $\T_q$ defined by (\ref{msa}).
\end{theo}\begin{demo}
It is enough to compute the limits of $\int \gl^n d\mu^A_{x}$
for $n$ fixed and $x \ra \infty$.
By definition, we have $\int t^n d\mu^A_{x}=[A^n](x,x)$,
 and  
\[[A^n](x,x)=\sum a_{x,x_1}a_{x_1,x_2}\cdots a_{x_{n-1},x}  ~\]
where the sum is on loops $\gg=(x,x_1,x_2,\cdots,x_{n-1},x)$ of length
$n$ based at  $x$.
 If we assume that $|x|_{\gG_0} >n/2$, the loops
do not meet $\gG_0$ and  therefore
  $[A^n](x,x)=[A_0^n](x,x)$.
\end{demo}

\subsection{The Fourier-Helgason transform}

Let us recall the definition of the  Fourier-Helgason transform
on the tree $\T_q$ with the root $O$. 
\begin{defi} \label{fht1}
For any $ f \in  C_0(\T_q)$, the  {\rm Fourier-Helgason transform}
 ${\cal F H}(f)$ is the function
defined by the finite sum  
\begin{equation} \label{fht}
 {\cal F H}(f)(\go,s):=\hat{ f}(\go,s)=
\sum_{x\in V_q}f(x)q^{(1/2 +is)b_{\go}(x)}~. 
\end{equation}
 for any $\go \in \Omega_O$
 and  any $s \in S$.\end{defi}

\begin{defi} \label{ipw} 
For any  $\go \in \Omega_O$ and any $s\in S$ we define the "incoming plane wave"
  $ e_0(\go,s)$ as the function $x\ra e_0(x;\go,s)$, where
\begin{equation*} 
 \forall x \in V_q, \quad e_0(x,\go,s) = q^{(1/2-is)b_{\go}(x)}.
 \end{equation*}
\end{defi} 

For $s\in S_0$, such a plane wave is a generalized eigenfunction
 for the adjacency
operator
 $A_0$ on $\T_q $ in the sense that it satisfies 
\begin{equation} \label{equ:fhti} 
  (\lambda_s-A_0)e_0(x,\go,s)=0\ ~ (\lambda_s=2\sqrt q  \cos (s\log q))~,
 \end{equation} 
but is not in $l^2$.

 If we restrict ourselves to $s \in S^0$,  definition \ref{fht1} writes
\begin{equation} \label{fhto}
 \hat{ f}(\go,s)= \langle  e_0(\go,s), f \rangle =\sum_{x\in V_\gG }
  f\left( x\right)\overline{ e_0(x,\go,s)}~,
\end{equation}  and the completeness of the set
 $\{e_0(\go,s),~s\in S^0,~ \go \in \gO\}$
 is expressed by 
the following inversion formula
 (see [CMS]):
\begin{theo}\label{fhti}
For any $f \in C_0(\T_q)$, the following inverse transform holds
\begin{equation} \label{fit}
 f(x)= \int_{S^0}\int_{\gO}e_0(x,\go,s)
 \hat{ f}(\go,s)d\gs_O(\go) d\mu (s)
 \end{equation}
where \begin{equation}\label{chgtit} 
 d\mu (s)=\frac{(q+1)\log q}{\pi} \frac{ \sin^2 (s \log q)}{
 q+q^{-1} -2 \cos (2s \log q)} |ds|~.
\end{equation}
Moreover the Fourier-Helgason transform extends
to a unitary map from $l^2(\T_q)$ into
$L^2 (\Omega \times S^0 ,d\sigma _O \otimes d\mu )$.

The Fourier-Helgason transform
is not surjective: its range is the subspace
 $L^2_{\rm even} (\Omega \times S^0 ,d\sigma _O \otimes d\mu )$
 of the functions $F$ of $L^2 (\Omega \times S^0 ,d\sigma _O \otimes d\mu )$
which satisfy the symmetry condition (see, for example, \cite{CS} or \cite{FN})
\begin{equation*} \label{symmcd}
\int_{\gO}e_0(x,\go,s)F(\go,s)d\gs_O(\go)=
 \int_{\gO}e_0(x,\go,-s)F(\go,-s)d\gs_O(\go)\ .
\end{equation*}
The Fourier-Helgason transform provides a spectral resolution of
$A_0$:
if $\phi :\R \ra \R$ is continuous,
\[ \phi (A_0)={\cal (FH)}^{-1} \phi(\gl_s){\cal FH}~,\]
where $ \phi(\gl_s)$ denotes the operator of multiplication by that
function on $L^2_{\rm even}(\gO \times S^0,d\gs_0 \otimes d\mu )$.
 \end{theo}

\begin{coro}
From the inverse  Fourier-Helgason transform formula
(\ref {fit}) we find back the expression of the spectral measure of
$\T_q
$ (see Theorem \ref{cdv}).
\end{coro}
\begin{demo}
By homogeneity of the tree $\T_q$, for any continuous function $\phi:\  \R \ra ~\R $,
 $[\phi(A_0)](x,x)$
is independent of $x$. Using 
(\ref {fit}), we get

\begin{equation*} 
 [\phi(A_0)](O,O) =\int_{\gO}\int_{S^0} \phi (\gl_s)
\overline{e_0(O,\go,s)} e_0(O,\go,s)\ d\gs_O(\go)  d\mu(s )\ =
 \int_{S^0} \phi (\gl_s)
\  d\mu(s ).
 \end{equation*}
Let us perform the change of variables \begin{equation}\label{chge}
s =f_q(\gl):=\frac{1}{\log
  q}\arccos
  \frac{\gl}{2\sqrt q}\ .\end{equation}
Using (\ref{chgtit}) and the fact that, by the map $s \ra \lambda_s$, 
the circle $S^0$ is a double covering of $I_q$, we write

\begin{equation*} 
 [\phi(A_0)](x,x) \ = 2 \frac{(q+1)\log q}{\pi}\int_{I_q}
 \phi (\gl) \frac{ 1- \gl^2/4q}{
 q+q^{-1} +2 -\gl^2/q}f'_q(\gl) d\gl 
\end{equation*}
\begin{equation*} 
 \quad\quad \ = 2 \frac{(q+1)}{4\pi}\int_{I_q}
 \frac{\sqrt{4q-\gl^2}}{(q+1)^2-\gl^2}  \phi (\gl) 
  d\gl \ ,
\end{equation*}
which actually implies formula (\ref{msa}).
\end{demo}
\section{A  scattering problem for a Schr\"odinger operator
with a compactly supported non local potential }
\label{sec:schro}

We are concerned here with  the scattering on 
 $\T_q$ between the adjacency operator $A_0 $
 and the Schr\"odinger operator 
 $A= A_0+W$, where $W$ is a compactly supported non local potential.
 More precisely the Hermitian  matrix (also denoted $W$) associated to
this potential is
supported by $K\times K$ where $K$ is
a finite part of $V_q$. We assume in what follows that $K$ is chosen
minimal, so that:
\[ K=\{ x\in V_q~|~\exists y \in V_q {\rm ~ with ~}W_{x,y}\ne 0 \}~.\]  
Let us first describe the spectral theory of $A$:
it follows from \cite{RS}, Sec. XI 3,
 and from the fact that $A$ is a finite rank
perturbation
of $A_0$  (see also Section \ref{ss:ppsp}) that the Hilbert space
 $l^2(\T_q)$ admits an orthogonal
decomposition into two  subspaces invariant by $A$:
$l^2(\T_q)={\cal H}_{\rm ac} \oplus {\cal H}_{\rm pp}$
where 
\begin{itemize}\item 
$ {\cal H}_{\rm ac} $ is the isometric image of $l^2(\T_q)$ by the
{\it wave operator}

 $$ \Omega ^+ =s-\lim_{ -\infty}e^{itA}e^{-itA_0}~.$$
We have $A_{| {\cal H}_{\rm ac}}=\Omega ^+ A_0 (\Omega ^+)^\star $, so
that
the corresponding part of the spectral decomposition  is isomorphic
 to  that of  $A_0$  which is an absolutely continuous spectrum 
on the  interval
$ I_q$.
\item The space  ${\cal H}_{\rm pp}$ is finite dimensional,
admits an orthonormal basis of  $l^2$ eigenfunctions associated
to  a finite set of eigenvalues, some of them may be embedded in the
continuous spectrum $I_q$.
\end{itemize}
We will denote by $P_{\rm ac}$ and $P_{\rm pp }$ the orthogonal
projections on both subspaces.

In order to make the spectral decomposition more explicit, we will
introduce suitable  generalized eigenfunctions of $A$.
These  generalized eigenfunctions
  are  particular solutions of
\begin{equation}\label{hal} 
 (\lambda_s-A)e(.,\go,s) = 0~,
 \end{equation} 
meaning not $l^2$ solutions, but only point-wise solutions.
 For the adjacency operator $A_0$, we have seen that
 these generalized eigenfunctions,
  called the  `` plane waves''
 are given by the $e_0(\go,s)$'s with 
  $s\in S^0$ 
and $\go \in \Omega _O$ (see definition \ref{ipw}) and give  the Fourier-Helgason transform
which is the spectral decomposition of $A_0$ (Theorem \ref{fhti}).

 We are going to prove a similar  eigenfunction expansion theorem
for $A$,  using 
 generalized eigenfunctions of   $A$.
 We will mainly adapt the presentation of \cite{RS}, Sec. 
XI.6, for Schr\"odinger operators in $\R^3$ (see also \cite{Ike}).
Our first goal is to build the generalized eigenfunctions
$x\ra e(x,\go,s)$ also denoted $e(\go,s)$. We will derive and solve
the  so-called {\it Lippmann-Schwinger equation}. This is an
integral equation  that $e(\go, s)$ will
satisfy.

\subsection{Formal derivation of the Lippmann-Schwinger equation}
Let us proceed first in a {\it formal way} by transferring the
functions $e_0(\go,s)$ 
by  the wave operator:  
if $ e(\go,s)$ is the image of  $e_0(\go,s)$
 by the wave operator $\gO^+$ in some sense (they are not in $l^2$!),
 then we should have
$\Di e_0(\go,s)\ ={\rm lim}_{t\rightarrow -\infty}
 e^{itA_0}e^{-itA}e(\go,s)$\\
 $\Di\quad\ ={\rm lim}_{t\rightarrow -\infty} [e(\go,s)- i\int_0^{t}
  e^{iuA_0}We^{-iuA}e(\go,s)du]$\\
$\Di\quad\ = e(\go,s)- i{\rm lim}_{\ge\rightarrow 0}\int_0^{-\infty}
  e^{iuA_0}We^{-iu\gl_s}e^{\ge u}e(\go,s)du$\\
$\Di\quad\ = e(\go,s)+ {\rm lim}_{\ge\rightarrow 0}[(A_0
-(\gl_s+i\ge))^{-1}
We(\go,s)]\ .$\\

So $ e(\go,s)$ should obey the following "Lippmann-Schwinger-type" equation

\begin{equation} \label{foitt}
e(\go,s) =e_0(\go,s)+
 G_0(\gl_s)W e(\go,s) ~.
 \end{equation}

\subsection{Existence and uniqueness  of the solution for the
 modified  "Lippmann-Schwinger-type" equation }
\label{sec:lsmodif}
Let $\chi \in C_0(\T_q)$ be a  compactly supported real-valued
 function so that $W\chi =\chi W=W$. For example $\chi $
can be the characteristic function of $K$.
We first introduce a  modified  "Lippmann-Schwinger-type" equation.
 If  $ e(\go,s)$ obeys (\ref{foitt}) and $a(\go,s)=
\chi  e(\go,s) $, then $a$ obeys
\begin{equation} \label{mlse} 
a(\go,s) = \chi e_0 (\go,s) +
\chi G_0(\gl_s)W a(\go,s)~.
\end{equation}

We have the following result~: 

\begin{prop}\label{indep} Let $\chi \in C_0(\T_q)$ be a  compactly supported real-valued
 function so that $W\chi =\chi W=W$. Set  \begin{equation}
\hat{\cal E}=:\{s \in S^0; \ker ({\rm Id}-\chi G_0(\gl_s)W)\ne 0\}
\end{equation}
\begin{enumerate}
\item The set $\hat{\cal E}$ 
 is finite and    independent
of the choice of $\chi $.
\item If $s \notin \hat{\cal E}$, then (\ref{mlse}) has a unique solution $a(\go,s) \in  C_0(\T_q)$ 
and the function $e(\go,s)=e_0(\go,s) +G_0(\gl_s)W a(\go,s)$
is the unique solution of  the Lippmann-Schwinger equation
(\ref{foitt}).

\item  The set
$\hat{\cal E}$ is invariant by  $s \ra -s$ and consequently
it is the pre-image by $ \tilde\gL: s\ra \gl_s$ of a subset of $I_q$ which we 
 denote by  ${\cal E}$.  \end{enumerate}
\end{prop}
\begin{demo}
We first prove 2). Let $L_{s,\chi}$ be the finite rank  operator on $l^2(\T_q)$ defined by
 $L_{s,\chi}=\chi G_0(\gl_s)W $. The map $s\ra L_{s,\chi}$  extends holomorphically
to $\Im s >-\ha $. Equation (\ref{mlse})
takes the form
\begin{equation} \label{mlseb} 
a(.,\go,s) = \eta(.,\go,s) +L_{s,\chi} a(.,\go,s),
\end{equation}
where $\eta(.,\go,s)\in C_0(\T_q)$. By
  the analytic Fredholm
theorem (\cite{RS}, p. 101),
 there exists a finite  subset $\hat{\cal E}$
 of $S^0 $, defined by $\hat{\cal E}=:\{s \in S^0; \ker ({\rm Id}-L_{s,\chi})\ne 0\}$,
so that 
equation (\ref{mlse}) has a unique solution $a(\go,s) \in  C_0(\T_q)$
 whenever $s \notin \hat{\cal E}\ .$ 
The second assertion of 2) comes from the fact that $W a(\go,s)=
W\chi  e(\go,s)=W e(\go,s)$.

Let us now prove  1):  
the ``minimal'' $\chi $ is $\chi _W ={\bf 1}_{\rm K}$.
If $a$ is a non trivial solution of $a-\chi_W G_0(\gl_s)Wa =0$,
and $\chi \chi _W= \chi _W$, $a$ is also solution of 
 $a-\chi G_0(\gl_s)Wa =0$.

Conversely, if  $a-\chi G_0(\gl_s)Wa =0$, we have 
  $\chi_W a-\chi_W G_0(\gl_s)W\chi_W a =0$. We have to prove that
$\chi_W a \ne 0$. If $\chi_Wa =0$,
we would have $Wa=0 $ and $a=0$.

To prove 3) it is enough to notice that for any $s\notin \hat{\cal E}$,
 we have $L_s=L_{-s}$.
\end{demo}

\subsection{The set ${\cal E}$ and the pure point spectrum}
\label{ss:ppsp}

\begin{prop}\label{compact}
If $(A- \gl )f=0$ with $\gl \in I_q$ and $f \in l^2(\T_q)$,
then ${\rm Supp}(f)\subset \hat{K}$ where $\hat{K}$ is the smallest subset of
$V_q$ so that ${\rm Supp}(W)\subset \hat{K}\times \hat{K} $
 and all connected components of $\T_q \setminus \hat{K}$ are infinite.
\end{prop}

\begin{demo} We will proceed by contradiction.
Let $x \in V_q\setminus \hat{K}$ be so that $f(x)\ne 0$.
Let us define an infinite  sub-tree $\T_x$ of $ \T_q$
as follows: let $y_\ga, \ga =1,\cdots,a $ be the
vertices of $\T_q$ which satisfy $y_\ga \sim x$ and
$y_\ga $ is closer to $\hat{K}$ than $x$.
Then $\T_x$ is the connected component of $x$ in the graph obtained
from $\T_q $ by removing the edges $\{ x,y_\ga \} $
for $\ga =1,\cdots, a$.
Let us consider the "averaged" function
 $$n \in \N \ra  \bar{f}_x(n):=\frac{1}{q^n}\sum_{z\in \T_x,~d(x,z)=n}
 f(z)\ .$$
 Then
$\bar{f}_x$ satisfies the ordinary difference
 equation $\gl g(n)-qg(n+1)-g(n-1)=0$.

 We thus get a contradiction, since this equation has no non-zero  $l^2$
 solution when $\gl $ is in $I_q$ and hence
$f(x)=\bar{f}_x (0)=0$. 
\end{demo}
 
\begin{figure}[hbtp]
\leavevmode \center
\input{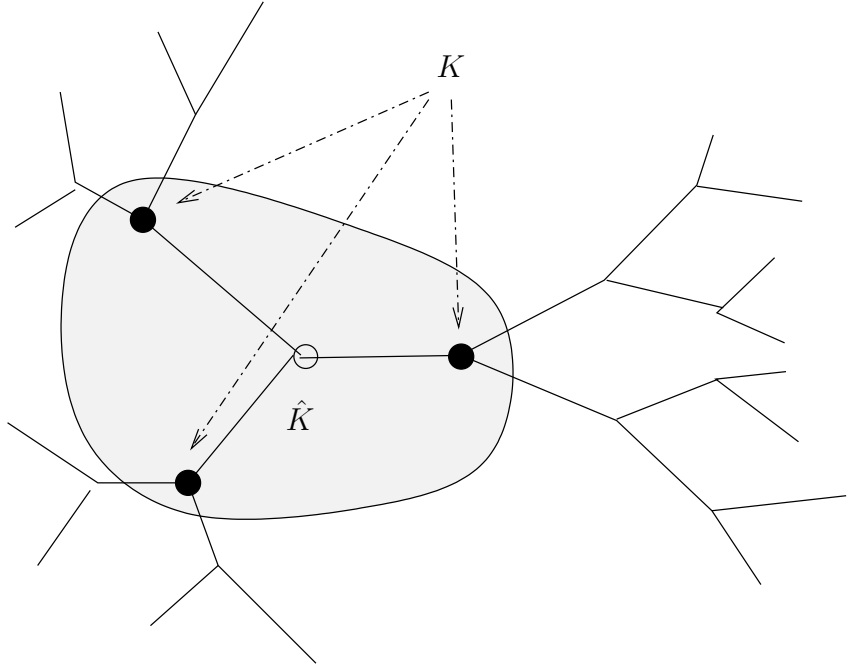}
\caption{{\it A simple example with $\hat{K}$ strictly larger than
$K$ }}
 \label{fig:hatk}
\end{figure}

\begin{coro} $\# \{ \gs_{\rm pp}(A) \cap I_q \} \leq \# \hat{K} $.
\end{coro}
This holds because any  eigenfunction associated to an eigenvalue
 in $\{ \gs_{\rm pp}(A) \cap I_q \}$ is supported in $\hat{K}$
and the dimension of the vector space of functions
supported in $  \hat{K}$
is $\# \hat{K}$.

\begin{theo} If $s \in S^0$, $(A-\gl_s )f=0$ and $f\in l^2(\T_q)\setminus 0$, 
then $s \in \hat{\cal E}$.

Conversely, if $s \in 
\hat{\cal E}\subset S^0$,  there exists $f\ne 0$
so that  $(A-\gl_s )f=0$ and $f(x)=O\left(q^{-|x|/2}\right)$.
\end{theo}
\begin{demo}Due to Proposition  \ref{compact},
 the support of such an $f$ is included in $\hat{K}$
and $(\gl_s-A_0)f=Wf$. We can apply $G_0(\gl_s)$ to both sides of the
equation, (although $\gl_s $ is in the spectrum of $A_0$)
because the functions on both sides are finitely supported.
 On the lefthandside we have $G_0(\gl_s )(\gl_s -A_0)f =f$:
 this is true for $\gl_s \notin I_q$
because $I_q$ is the spectrum of $A_0$ and hence by continuity
 ($G_0(\gl_s)$ extends holomorphically near $S^0$)
for every $\gl _s $ since  $f$ is compactly supported. 
Hence applying $G_0(\gl_s)$ to both sides yields $f=G_0(\gl_s)Wf $.
 Due to proposition \ref{indep}, we can choose for $\chi $ the characteristic
function of $\hat{K}$, so we get
$f- \chi G_0(\gl_s)Wf=0$. We have a non trivial solution of
$a-L_sa=0$, namely $a=f$.

Conversely, let us start from $a$, a non trivial solution of
$a-L_sa=0$ and define $f=G_0(\gl_s)Wa $.
Then
 \[ (\gl_s-A)f=(\gl_s-A_0 )G_0(\gl_s)Wa -WG_0(\gl_s)Wa \]
and $(\gl_s-A_0 )G_0(\gl_s)Wa = Wa $ by analytic extension from $s \in
S^+$.
Hence, using $ W \chi =W$,
 \[ (\gl_s-A)f=Wa - W \chi G_0(\gl_s)Wa  =Wa - W a=0 ~.\] 
 From the definition of $f$, we get that  
$f$ is a finite linear combination of the functions  $G_0(\gl_s,.,y),
~y\in {\rm Supp}(W)$ and we can use Equation (\ref{rad}) to get
 the bound in $x$.

\end{demo}
\begin{theo} The pure point spectrum $\gs_{\rm pp}(A)$ of $A$ splits into 
3 parts 
\[  \gs_{\rm pp}(A)=  \gs_{\rm pp}^-(A) \cup  \gs_{\rm pp}^+(A)
\cup   \gs_{\rm pp}^0(A) \]
where $ \gs_{\rm pp}^-(A)= \gs_{\rm pp}(A)\cap ]-\infty ,-2\sqrt{q}[$,
  $ \gs_{\rm pp}^+(A)= \gs_{\rm pp}(A)\cap ]2\sqrt{q}, +\infty [$,
and  $ \gs_{\rm pp}^0(A)= \gs_{\rm pp}(A)\cap I_q$.
We have $\#  \gs_{\rm pp}^\pm (A)\leq \# {\rm Supp}(W)$
and  $\#  \gs_{\rm pp}^0 (A)\leq \# \hat{K}$.
\end{theo}
The first  estimate comes from the mini-max principle and the fact
that $W$ is a rank $N$ perturbation of $A_0$ with $N= \# {\rm
  Supp}(W)$.
The second  one is already proved.

The reader could ask if there can really be some compactly
supported eigenfunctions. They can exist as shown by the following
2 examples. 
\begin{exem}
$\Gamma $ is a tree with root $O$ and $W_{x,0}=W_{0,x}=-1 $ for 
any $x\sim O$. All other entries of $W$ vanish. Then if $H=A_\Gamma
+W$,
$f=\gd (0)$, we have $Hf =0$.
\end{exem}
\begin{exem}
The graph $\Gamma $ is the union of a cycle with 4 vertices $\{
1,2,3,4 \}$  and
a tree whose root is attached to 2 neighboring vertices of the cycle.
If $f(p)=(-1)^p $  on the cycle and $0$ on all other vertices,
$A_\Gamma f=0$.
\end{exem}
However the proof of the following result is left to the reader:
\begin{prop} If $\Gamma $ is an infinite tree, then $A_\Gamma $
has no compactly supported eigenfunction.
\end{prop}

\subsection{The deformed Fourier-Helgason transform}
\label{ss:defFH}
\begin{defi} \label{fht2}
We define the deformed Fourier-Helgason transform ${\cal FH}_{\rm sc}$
of $f\in C_0(\T_q)$ as the function $\hat{ f}_{\rm sc}$
on $\Omega \times (S^0\setminus \hat{\cal E})$ defined by 
\begin{equation} \label{gfht}
 \hat{ f}_{\rm sc}(\go,s)= \langle  e(\go,s), f \rangle =\sum_{x\in V_\gG }
  f\left( x\right)\overline{ e(x,\go,s)}~.
\end{equation}
\end{defi}

We want to prove the following
\begin{theo}\label{fhtil}
For any $f \in C_0(\T_q)$ and any closed interval 
$J\subset I_q \setminus \cal E$, if we denote
by ${\hat{J}}$ the inverse image of $J$ by $s\ra \gl_s$, 
 the following inverse transform holds
\begin{equation} \label{fhi}
 P_{J}f(x)= \int_{{\hat{J}}}\int_{\gO} e(x,\go,s)
 \hat{ f}_{\rm sc}(\go,s)d\gs_O(\go)d\mu (s)~.
 \end{equation}
Moreover, $f\ra \hat{f}_{\rm sc}$ extends to an isometry
from ${\cal H}_{\rm ac}$ onto $L^2_{\rm even} (\Omega \times S^0,d\sigma
_O\otimes d\mu) $.
  \end{theo}

\subsubsection{ The relation of the deformed Fourier-Helgason
 transform with  the resolvent}

Denoting, with a slight abuse of notation,
 for $s\in S^+$,   by $ G(s)$  the operator
$(\lambda_s-A)^{-1}$
and similarly by $G_0(s)$ the operator $(\lambda_s-A_0)^{-1}$, 
 we have the  resolvent equation
\begin{equation}\label{resol}
G(s)=G_0(s)+G_0(s) W G(s)
\end{equation}

For $\gs \in S^0 $ and $s$ in $S^+$, we set
\begin{equation*}\label{resolinv}
h(s; \go,\gs ) = (\gl_s- \gl_{\gs}) G(s)e_0(\go,\gs)~,
\end{equation*}
where the right hand side  is a convergent series
which identifies to  $ (\gl_s- \gl_{\gs})$-times the inverse Fourier-Helgason
transform of $y \ra G(s; x,y)$. 

Using the definition of $G_0$ and (\ref{equ:fhti}) we have
\begin{equation*}\label{resoo}
(\gl_s- \gl_{\gs}) G_0(s)e_0(\go,\gs) = e_0(\go, \gs)+ A_0 G_0(s)e_0(\go,\gs) -G_0(s)[\gl_{\gs} e_0(\go,\gs)]= e_0(\go, \gs)\ ,
\end{equation*}
so equation (\ref{resol}) for $G$
 gives  an integral equation for $h$

\begin{equation*}\label{resoliin}
h(s;\go, \gs) = e_0(\go, \gs)+ G_0(s)W h(s;\go, \gs)
\end{equation*}
and, if $p(s;\go, \gs)=\chi h(s; \go, \gs)$,
\begin{equation} \label{mlsei} 
p(s;\go, \gs) = \chi e_0 (\go, \gs)+\chi G_0(s)W p(s;\go, \gs)\ .
\end{equation}
The key fact is the relation between (\ref{mlsei}) and
 the modified "Lippmann-Schwinger-type" equation (\ref{mlse}).
If $s \in S^+ $ is fixed and $\gs=s$, then the equation
 for $p(s;\go, s)$ is identical to equation (\ref{mlse})
 for $a(\go, s).$

This can be used to prove  
\begin{lemm}\label{endd}
Let us consider $f\in C_0(\T_q) $ , $\go \in \gO$ and $s\in S^+$. Then the function
\begin{equation*}\label{func}
\gF(s;\go, \gs) = \sum_{x\in V_q}\overline {h(x;s; \go, \gs)}f(x) , \quad \quad \forall \gs \in S^0
\end{equation*}
has a holomorphic
extension in $\gs$ to    $S^+$ and
$$\gF(s;\go, s) = \sum_{x\in V_q}\overline{e(x,\go,s)}f(x)=\hat{ f}_{\rm sc}(\go,s)\ .$$
\end{lemm}
We thus have related $\hat{ f}_{\rm sc}$ to  the resolvent.

\subsubsection{End of the proof of Theorem \ref{fhtil}}
Let $\gl_s= \gL+i\ge$
with $\gL \in I_q \setminus \cal E$ and 
 $\ge>0$, and $s\in S^+$ (this implies $0< \Re s < \tau/2$).
 Up to a factor of
 $(\gL+i\ge- \gl_{\gs})$, $(\go, \gs )\ra h(x;s;\go,\gs )$ is the inverse
 Fourier-Helgason transform of
$y\ra G(\gl_s,x,y)$;  so the Plancherel theorem implies (after multiplying
 by $\overline{f(x)}f(y)$) that
\begin{equation*}\label{resolin}
(\gl_s-{\overline\gl_s}) \sum_{z\in V_q}\overline{G({\overline\gl_s},x,z)}
G({\overline\gl_s},z,y)\overline{f(x)}f(y)=...\end{equation*}
\begin{equation*}....2i\ge 
\int_{S^0}\int_{\gO}\frac{ h(x;s;\go, \gs) \overline{h(y;s;\go, \gs)}
\overline{f(x)}f(y)}{|\gl_{\gs}-\gL|^2 +\ge^2 }d\gs_O(\go) d\mu(\gs)
\end{equation*}
If we sum over all $x$'s and $y$'s,
 we obtain for the left-hand side
\begin{equation*}
(\gl_{s}-\overline{\gl_s})\langle G(\overline{\gl_s})f,
G(\overline{\gl_s})f \rangle=(\gl_{s}-\overline{\gl_s})\langle 
 f| G(\gl_s) G(\overline{\gl_s}) f  \rangle=\langle f|
[G(\lambda_s)-G({\overline\lambda_s})]f   \rangle
\end{equation*}
whereas the right-hand side becomes 
\begin{equation*}
\int_{S^0}\int_{\gO}\frac{ 2i\ge }{|\gl_{\gs}-\gL|^2 +\ge^2|}|
\gF(s; \go, \gs)  |^2d\gs_O(\go)  d\mu(\gs)\ .
\end{equation*} 
We thus conclude that, for any closed sub-interval $J$ of $I_q$ disjoint
from $\cal E$, 
\begin{equation*}\label{theend}
\frac{1}{2\pi i} \int_{J}  \langle f |[G(\gL+i\ge)-G(\gL-i\ge)]f 
 \rangle d\gL  =\frac{ 1}{\pi } \int_{J}  d\gL 
\int_{S^0}\int_{\gO}\frac{ \ge }{|\gl_{\gs}-\gL|^2 +\ge^2}|
\gF(s, \go,\gs)  |^2d\gs_O(\go) d\mu(\gs) 
\end{equation*}

%d\gs_0 devient d\gs_O
% rajouté un [

 As $\ge\rightarrow 0$, Stone's formula  implies that the left-hand
 side
 approaches $\|P_{J} f\|^2$. Moreover the measures 
$$dl_\ge = \frac{ \ge d\gL  }{\pi[|\gl_{\gs}-\gL|^2 +\ge^2]}$$
converge weakly to $\gd (\gL -\gl_{\gs})$
as $\ge \ra 0^+$. So one has to put $\gL =\gl_\gs$, which
implies $\gs=\pm s$.

Thus one gets that the right-hand side tends to
$\int_{\hat{J}}\int_{\gO}
| \hat{ f}_{\rm sc}(\go,\gs)|^2 |d\gs_O(\go)  d\mu(\gs )$,
where $\hat{J}$ is
 the inverse image of $J$
by $s\ra \gl_s$.

\section{The spectral theory for 
a graph asymptotic to a regular  tree}\label{sec:combi}
We are concerned here with  the spectral theory of the adjacency
matrix of  a graph $\gG$  
asymptotic
 to a regular  tree  of degree $q+1$,  in the sense of Definition
\ref{astl}, which we recall here: 
\begin{defi} 
Let $q\geq 2$ be a fixed integer. We say that the infinite graph $\gG$ is
  {\rm  asymptotic to a regular  tree of degree $q+1$}
 if $\gG $ is connected 
and  there exists a finite connected sub-graph $\gG_0$ of $\gG$  such that 
$\gG':=\gG\setminus  \gG_0$ is a disjoint union of a finite number
of trees $T_l,~l=1,\cdots, L,$ rooted at a vertex $x_l$ linked
 to $\gG_0$ and so that
all vertices of $T_l$ different from  $x_l$ are of degree $q+1$. 
The trees $T_l,~l=1,\cdots, L,$ are called the {\rm ends} of $\gG$. 
\end{defi}

  We want to reduce the spectral theory of $A_\gG $ to
the situation studied in Section \ref{sec:schro}.
For that, we need a preliminary combinatorial study which
could be of independent interest.

\subsection{Some combinatorics}

We need the following combinatorial result:
\begin{theo}\label{theo:comb}
If $\gG $ is  asymptotic to a regular  tree of degree $q+1$,
then $\gG $ is isomorphic  to a connected component 
 of a graph $\hat{\gG } $ which can be obtained from $\T_{q}$
by adding and removing  a finite number of edges.
\end{theo}
\begin{rem} By removing a finite number of edges, one could assume that
$\gG $ is a tree. Then the result is quite elementary if
the degree of all vertices of $\gG$  is $\leq q + 1$: it is then enough
to add infinite regular trees to the vertices of degrees $<q+1$
in order to get the final result. This argument, suggested by the
referee,
is not enough to give a complete proof. \end{rem}

In order to prove Theorem \ref{theo:comb}, we first introduce
an integer  $\nu (\gG) $ associated to the graph  $\gG $;
the integer $\nu $ is a combinatorial analogue of the
regularized  total curvature
of a Riemannian surface $S$  which is of constant  curvature
$\equiv K_0 $  near  infinity,
namely $\int _S (K-K_0) |d\gs | $.
\begin{defi}If $\gG $ is  asymptotic to a regular  tree of degree
  $q+1$,
we define $\nu (\gG )$ by 
\[ \nu (\gG)=\sum _{x\in V_\gG} (q+1 -d(x) )+2 b_1 ~,\]
where $d(x)$ is the degree of the vertex $x$
and $b_1 $ is the first Betti number of $\gG$  or equivalently the number
of edges to be removed from $\gG $ in order to get a tree.
\end{defi}
Note that, if $T$ is a maximal sub-tree of $\gG $,
$\nu(T)=\nu(\gG)$.

We will need the
\begin{lemm}\label{lemm:nu}
 If, for $r\geq 2 $,  $B_r=\{ x\in V_\gG ~|~ |x|_{\gG_0}
 \leq r \}$, then
we have
\[ \nu (\gG)= (q-1)m -M +2 ~,\]
where $m $ is the number of inner vertices of $B_r$
and $M$ the number of boundary vertices (i.e. connected to
a vertex  of $\gG \setminus B_r $) of $B_r$. 
\end{lemm}
\preuve 
Each of the $M$ boundary vertices has $q$ neighbors in  $\gG \setminus B_r $
and one in $B_r$.
 From Euler formula applied to the sub-graph $\gG  \cap B_r$ which is
 connected
by the assumption on $\gG_0$,
we get 
\[ 1-b_1= (m+M) - \ha \left( \sum _{|x|_{\gG_0}\leq r-1} d(x)
 +M \right)~.\]
Thus  
\[ \nu (\gG )= \sum _{|x|_{\gG_0}\leq r-1} (q+1-d(x)) +2b_1\]
is equal to
\[ \nu (\gG)= (q+1)m -(2m +M -2+ 2b_1)+2b_1~.\]
\finpreuve

We will also need the:
\begin{lemm}\label{lemm:counting}
 Let  $F $ be  a finite  tree whose all vertices are of degree $q+1$
except the ends which are of degree $1$. 
Let   $M$ be the  number of ends and $m$  the number of inner
vertices of $F$.
We have the relation  
\begin{equation}\label{equ:euler}
M=2+(q-1)m ~.\end{equation}

Conversely, for each choice of $(m,M)$ satisfying 
Equation (\ref{equ:euler}), there exists such a tree $F$.
\end{lemm}
\preuve From Euler formula applied to $F$, we get
$1=|V_F| -|E_F|$. Moreover 
$|V_F|=m+M$. Let us  choose  a root inside $F$
and orient the edges from that root.
Then we count the edges by partitioning them 
with their $m$ possible origins; this gives  $|E_F| =(q+1)+ (m-1)q $.

Conversely, the statement is true for $m=1,~M=q+1$
and we proceed by induction on $m$ by adding $q$ edges
to a boundary vertex and the corresponding $q$ boundary vertices, we
have $m\ra m+1,~M\ra M+(q-1)$. 
\finpreuve
\begin{lemm}\label{lemm:nu=0} If $\gG $ is asymptotic to a regular tree
of degree $q+1$, $\gG $ can be obtained from a tree
$\T_q$ by removing and adding a finite number of edges 
if and  only if $\nu (\gG )=0$.
\end{lemm} 
\begin{figure}[hbtp]
\leavevmode \center
\input{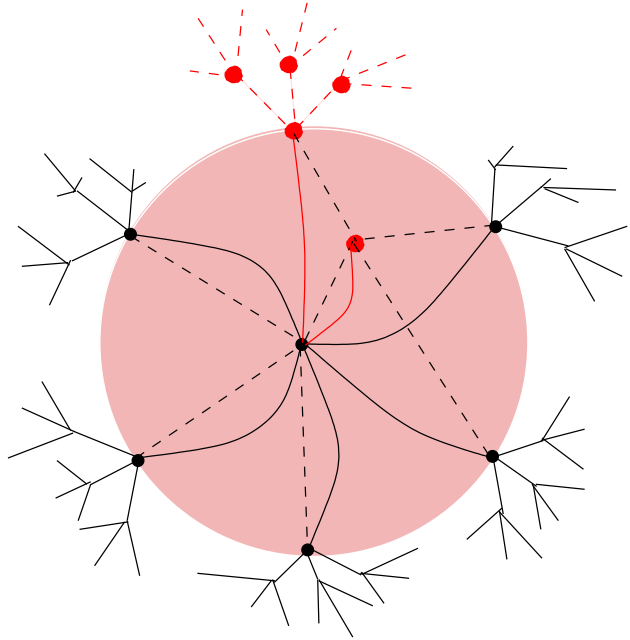}
\caption{{\it \small  Changing the graph with $\nu =0$ into $\T_q$:
the dashed edges are the new edges, the continuous one the old edges.
The picture is done in the same situation as in Figure \ref{fig:theo41}.}}
 \label{fig:treeF}
\end{figure}
\preuve
All the changes will take place inside
the sub-graph $B_r$. 
If we denote by $M$ the number of boundary vertices
and $m$ the number of inner vertices of $B_r$, we have, using 
$\nu (\gG )=0$ and  Lemma \ref{lemm:nu}, 
$ M=2+(q-1)m$.  
We replace the graph $B_r$ by a tree $F$ whose existence is stated
in Lemma \ref{lemm:counting}. The vertices of both graphs are the same
and all vertices of the new graph have degree $q+1$.
Hence the  new graph is a regular tree $\T_q$.
\finpreuve

We will now make some modifications of $\gG $ in order to 
get a new graph $\hat{\gG }$ with $\nu (\hat{\gG }  )=0$.

\begin{lemm}\label{lemm:moves}
 If $\gG'=M_1(\gG )$ is defined by adding to $\gG $ a vertex and an
  edge
connecting that vertex to a vertex of $\gG_0 $, then
$\nu(\gG ')=\nu(\gG )+q-1 $.

If  $\gG ''=M_2(\gG ) $ is defined by adding
 to $\gG $ a tree whose root $x$ is of
degree
$q$ and all other vertices of degree $q+1$ and connecting $x$ by an
edge
to a vertex of $\gG _0 $, $\gG ''$ is asymptotic to a regular  tree
of degree $q+1$ and $\nu(\gG ')=\nu(\gG )-1 $.
\end{lemm}
This Lemma is quite easy to check.

\begin{figure}[hbtp]
\leavevmode \center
\input{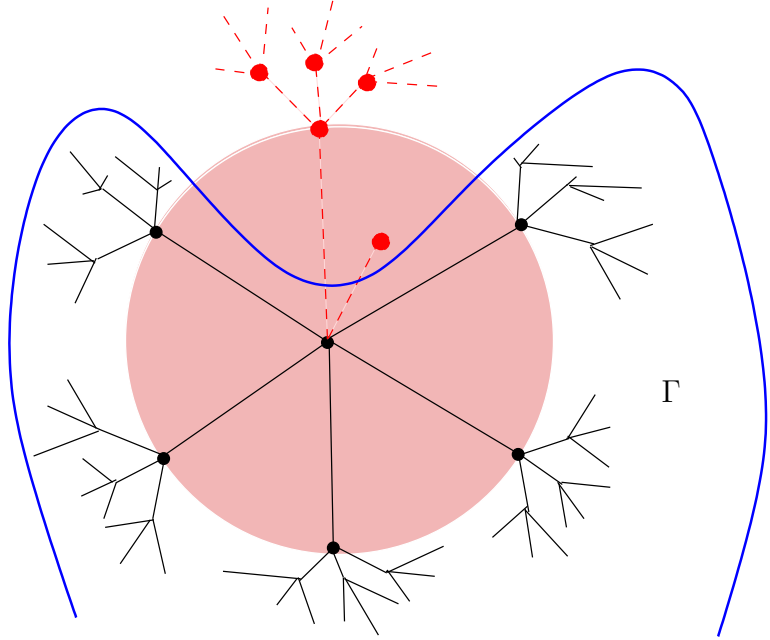}
\caption{{\it \small The construction in the proof of Theorem \ref{theo:comb}; 
the graph $\gG $ has $q=3$, $\nu =-1$, $N'=N''=1$, $m=2$, $M=6$.}}
 \label{fig:theo41}
\end{figure}

\noindent {\it Proof of Theorem \ref{theo:comb}.--~}
Let us now write
$\nu(\gG)= N''-(q-1)N'$ with $N'\geq 0 $
and $N''\geq 0$. By performing $N'$ times the move $M_1$
and $N''$ times the move $M_2$, we arrive to 
a graph $\tilde{\gG}$ with $\nu (\tilde{\gG})=0$.
Let $\hat{\gG}$ be the graph obtained by removing
from $\tilde{\gG}$ the $(N'+N'') $  edges not in $E_\gG $,
 one of whose vertices is in
$\gG _0$. The graph $\hat{\gG }$
is clearly asymptotic to a regular tree
of degree $(q+1)$ and  $\gG $ is a connected component of $\hat{\gG}$.

It remains to prove that, by removing and adding a finite 
number of edges to   $\hat{\gG}$, we get a tree $\T_q$:
this is the content of Lemma \ref{lemm:nu=0}.
\finpreuve

\subsection{The spectral theory of $\gG $}

From Theorem \ref{theo:comb}, we can identify the set of vertices
of $\gG $ to a subset of the set of vertices of $\hat{\gG}$ which
is the same as the set of vertices of $\T_q$.
 We deduce the existence
of  a Hilbert space $ {\cal H} $ so that
$l^2(\T_q)=l^2 (\gG )\oplus {\cal H} $ and this decomposition
is invariant by $A_{\hat{\gG}}$. Moreover
 $A_{\hat{\gG}}$ is a finite rank perturbation of $ A_0=A_{\T_q}$.
This will allow us to describe the spectral theory of
$A_\gG $ by using the results of Section  \ref{sec:schro}.

In order to get the spectral decomposition of
$A_\gG $ in terms of the spectral decomposition
of $A_{\hat{\gG}}$ given in Section \ref{ss:defFH}, we will need the
\begin{lemm} \label{lemm:support}
Let $A_{\hat{\gG}} = A_{\T_q} +W$ with ${\rm Support}(W) \subset
K\times K $ and $K $ finite.
Let $\gG $ be an unbounded  connected component of $\hat{\gG} $ and 
$\go $ a point at infinity  of $\gG $.
Then, for  any $s\notin \hat{\cal E}$,
we have
\[ {\rm support}(e(.,s,\go))\subset V_\gG ~.\]

Conversely, if $\go'$ is a point at infinity of $\hat{\gG }$
which is not a point at infinity of $\gG $ then 
\[ {\rm support}(e(.,s,\go'))\cap  V_\gG =\emptyset  ~.\]
\end{lemm}
\preuve 
We will  apply the general Theorem \ref{fhtil} in our combinatorial context.
Let us prove the first assertion, the proof of the second is similar.
It is enough to prove it for $s \in S^+$ close to $S_0$ and 
hence $\gl_s $ not in the spectrum of $A_0$,
 because $s\ra e(x,s,\go)$
is meromorphic on $S$.
We have then (Equation (\ref{foitt})) 
\[ e(s,\go)=e_0(s,\go) + G_0 (\gl_s)W e(s,\go)~.\]
From the explicit expression of
$e_0$ (see Definition \ref{ipw}), we get that the first term belongs to
 $l^2({\hat{\gG }}\setminus \gG ) $,
and so does the second one, as the image of a compactly supported function by
the resolvent  for $\gl_s \notin I_q={\rm spectrum}(A_0 )$
(recall that the resolvent is continuous in $l^2$).

This proves that the restriction   of $e(.,s,\go) $
to $V_{\hat{\gG }}\setminus V_\gG $ is an $l^2$ eigenfunction, with
eigenvalue
$\gl_s$, 
of $A_{\hat{\gG}}$. Since $A_{\hat{\gG}}$ has no eigenvalue $\gl_s$ for
$s\in S^+ $ close to $S_0$, it follows that 
$e(x,s,\go)$ vanishes for $s\in S^+$ close to $S_0$ and  $x\notin V_\gG $. 
\finpreuve

Theorem \ref{theo:comb} allows to consider the set $\Omega $  of
points at
infinity of $\Gamma $ as a subset  of the set $\hat{\Omega}$
of the  points at infinity of
$\hat{\Gamma}$. The space $l^2({\hat{\Gamma}})$ splits as a direct sum
$l^2(\Gamma)\oplus l^2({\hat{\Gamma}}\setminus \Gamma )$
which is preserved by the adjacency matrix. Lemma \ref{lemm:support}
shows that the support of the  generalized eigenfunctions
$e(.,s,\go)$ for $\go \in \Omega $ is included
in $V_\Gamma$.
Using this, we can state the spectral decomposition of
$A_\Gamma $ as an immediate corollary of Theorem \ref{fhtil}.     
\begin{theo}
The Hilbert space $l^2(\Gamma )$ splits into a finite dimensional
part ${\cal H}_{\rm pp }$ and an absolutely continuous part
${\cal H}_{\rm ac }$. This decomposition is preserved by $A_\Gamma $.
If $f\in C_0(\Gamma)$ and, for $\go \in \Omega $,
$\hat{f}(s,\go)=\langle e(.,s,\go),f \rangle $,
then the map $f\ra \hat{f}$ extends
to an isometry from ${\cal H}_{\rm ac }$ onto
$L^2_{\rm even}(S_0\times \Omega , d\gs_0 \otimes d\mu ) $
which intertwines the action of $A_\gamma $ with the multiplication
by $\gl_s$.
\end{theo}

\section{ Other features of the scattering theory in the setting of section 3}

We are again concerned here with  the scattering theory on 
 $\T_q$ between the adjacency operator $A_0 $
 and the Schr\"odinger operator 
 $A= A_0+W$, where $W$ is a compactly supported non local potential.
Let us recall that 
\[ K=\{ x\in V_q~|~\exists y \in V_q {\rm ~ with ~}W_{x,y}\ne 0 \}~.\]

\subsection{Correlation of scattered plane waves}

In the paper \cite{CdV3}, the first author computed the point-point
 correlations 
of the plane waves for a scattering problem in $\R^d$ in terms of the
Green's
function:
for a fixed spectral parameter, plane waves are viewed as random
waves
parametrised by the direction of their incoming part.
The motivation comes from passive imaging in seismology, a method
developped by Michel Campillo's seismology group in Grenoble, 
as described for example in the papers \cite{CdV4,CdV5}.
Following a similar method, we will compute the correlation
of plane waves for our graphs viewed as random waves parametrised by
points at infinity. 

From Theorem
\ref{fhtil} we get, for any $\phi \in C_0(\T_q)$ such that ${\rm supp}\ \phi \in I_q \setminus \cal E$, the 
following formula for the kernel of  $\phi(A)$:

\begin{equation*} 
 [\phi(A)]_{x,y} =\int_{\gO}\int_{S^0 \setminus \hat{\cal E}} \Phi (\gl_s)
\overline{e_(x,\go,s)} e_(y,\go,s)\ d\gs_O(\go)  d\mu(s )
 \end{equation*}
Taking $\phi = 1_I$, the characteristic function of some 
 interval $I= [a, \gl]\subset I_q \setminus \cal E$,  we get:
  \begin{equation*} \label{chart}
 [\Pi]_I (x,y) = 2\int_{\gO}\int_{f_q(\gl)}^{f_q(a)}
 \overline{e(x,\go,s)}
 e(y,\go,s)\ d\gs_O(\go)  d\mu(s )
 \end{equation*}
where we set $f_q(t) =\frac{1}{\log q} {\rm Arccos} \frac{t}{2\sqrt q}$
 as in (\ref{chge}), and where we use  the fact that, by the map $s \ra \lambda_s$, 
the circle $S^0$ is a double covering of $I_q$. In particular $f_q(\gl_s)=s$.
In the sequel we  note $f_q(\gl)= s(\gl) $ for simplicity.

If we consider the plane wave $e(x,\go, s(\gl))$ for $\gl \in I_q \setminus\cal E$, as a random wave,
 we can define the point-to-point correlation $C_{\gl}^{sc}(x,y)$ of 
such a random wave in the usual way:
\begin{defi} For any $\gl \in I_q \setminus \cal E$, the point-to-point correlation $C_{\gl}^{sc}(x,y)$ of
 the random wave $e(x,\go, s(\gl))$ is given by
\begin{equation*} 
  C_{\gl}^{sc}(x,y) =  \int_{\gO} \overline{e(x,\go, s(\gl))}
 e(y,\go, s(\gl))\ d\gs_O(\go)\ .
 \end{equation*}
\end{defi}

Denoting again by $G$ l the Green's function of $A$:
 $(\lambda-A)^{-1}[x,y]:= G(\gl,x,y)$ for $Im \gl >0$ 
we prove the

\begin{theo}
For any $\gl \in I_q \setminus \cal E$ and any vertices $x,y $ the  point-to-point
 correlation can be expressed in terms of the Green's function as 
\begin{equation*} 
  C_{\gl}^{sc}(x,y) =  - \frac{ 2 (q^2+2q +1-\gl^2 ) }{(q+1)\sqrt{4q-\gl^2} }
\ \Im  G(\gl+i0,x,y)\ .
 \end{equation*}
\end{theo}
\begin{demo}
Taking the derivative with respect to $\gl$ in equation (\ref{chart}) yields:
\begin{eqnarray}
& \frac{d}{d\gl}&[\Pi]_I (x,y) =...\nonumber\\  
&=& -2 f'_q(\gl) \frac{(q+1)\log q}{\pi} \frac{ \sin^2 (s(\gl) \log q)}{
 q+q^{-1} -2 \cos (2s(\gl) \log q)}\ \int_{\gO}
\overline{e(x,\go,s(\gl))}
 e(y,\go,s(\gl))\ d\gs_O(\go)\nonumber\\
 &=& \frac{q+1}{2\pi}\frac{ \sqrt{4q-\gl^2} }{ (q^2+2q +1-\gl^2 ) }
\ \int_{\gO} \overline{e(x,\go,s(\gl))} e(y,\go,s(\gl))\ d\gs_O(\go) .
\nonumber \end{eqnarray}
Thus we have
\begin{equation*} \label{derpiii}
 \frac{d}{d\gl} [\Pi_I] (x,y) = \frac{q+1}{2\pi}
\frac{ \sqrt{4q-\gl^2} }{ (q^2+2q +1-\gl^2 ) }\   C_{\gl}^{sc}(x,y)\ .
 \end{equation*}
Now we use the resolvent kernel of $A$ : $(\lambda-A)^{-1}[x,y]:=
 G(\gl,x,y)$ for $Im \gl >0$ and  Stone formula (\ref{sff})
to write \begin{equation*} \label{sfpiii}
 [\Pi]_I (x,y) = -\frac{1}{\pi}  \int_{a}^{\gl}
 \Im  G(t+i0,x,y) dt 
 \end{equation*}
and get the result. 
\end{demo}
\subsection{The T- matrix and the S-matrix}

The Lippmann-Schwinger eigenfunctions $e(x,\go,s)$ are especially useful 
to describe the so-called $S-$matrix ($S= (\Omega^-)^*\Omega^+$).
 First we introduce the
following object: 
\begin{defi} \label{tmat}
Let $(\go,s)$ and $(\go',s')$ be in $\Omega \times (S^0\setminus \hat{\cal E})$.
Define
 
\begin{equation*} \label{tmatr}
 T(\go,s;\go',s') =\langle  W e_0(\go,s),e(\go',s')\rangle  =
\sum_{(x,y)\in V_q\times V_q}e(x,\go',s')  \overline{W(x,y)}
 \overline{ e_0(y,\go,s)}  \ .
 \end{equation*}
$T(.;.)$ is called the $T-$matrix.
\end{defi}
The goal of  this section is to establish a relation between $S$ and $T$ (Theorem \ref{tvers}). To get the result we will need   the following 
\begin{lemm}\label{ffff}
For any $f \in C_0(\T_q)$ 
\begin{equation}\label{ome}
{\cal FH}_{\rm sc}(\Omega^+ f)(\go,s)=\hat{ f}(\go,s)
\end{equation}
\end{lemm}
\begin{demo}
Suppose that we can prove
$\Di
{\cal FH}((\Omega^+)^* f)={\cal FH}_{\rm sc}(f) (=\hat{ f}_{\rm sc}) ,
$
then (\ref{ome}) follows from
$\Di
{\cal FH}_{\rm sc}(\Omega^+ f)={\cal FH}((\Omega^+)^*\Omega^+ f)=\hat{ f}\ .
$
So, by Plancherel formula it is enough to prove that
\begin{equation}
(f,\Omega^+ g)=\int_{S^0 \times \gO}\overline{\hat{f}_{\rm sc}
(\go,s)}\hat{ g}(\go,s)d\gs_O(\go) d\mu (s) \ .
 \end{equation}
In the sequel, we set $d\gS:= d\gs_O(\go)
 d\mu (s) $ to simplify notations.

We have   $\Di (f,\Omega^+ g)-(f,g)= i{\rm lim}_{\ge\rightarrow 0}
\int_{-\infty}^{0}e^{\ge t}(f,e^{itA}We^{-itA_0} g) dt\ .
$
But $$(f,e^{itA} h)=\int_{S^0 \times\gO}\overline{\hat{f}_{\rm sc}
(\go,s)}\ e^{i\gl_{s} t}\ \hat{h}_{\rm sc} (\go,s)\ d\gS
$$
if either $f$ or $h$ is in ${\cal H}_{ac}$.
As a result (using definition \ref{fht2})
$$(f,e^{itA}We^{-itA_0} g)= \sum_{x\in V_q}\int_{S^0 \times\gO}
\overline{\hat{f}_{\rm sc}(\go,s)}\ e^{i\gl_{s}t}\ 
(We^{-itA_0}g)(x)\overline{ e(x,\go,s)} d\gS . 
$$
Thus$${\rm lim}_{\ge\rightarrow 0}\int_{-\infty}^{0}e^{\ge t}
(f,e^{itA}We^{-itA_0} g) dt\quad\quad\quad\quad\quad\quad\quad\quad\quad\quad\quad\quad$$
$$\quad ={\rm lim}_{\ge\rightarrow 0}\sum_{x\in V_q}
\int_{-\infty}^{0}\int_{S^0 \times\gO}\overline{\hat{f}_{\rm sc}(\go,s)}\ W(x)\ 
e^{-it(A_0 -\gl_{s}+i\ge )}\ g(x)\overline{ e(x,\go,s)}\  d\gS
$$
$$\quad =-i {\rm lim}_{\ge\rightarrow 0}\sum_{x\in V_q}\int_{S^0 \times
\gO}\overline{\hat{f}_{\rm sc}(\go,s)}\ W (x)\ 
[(A_0 -\gl_{s}+i\ge )^{-1}g](x) \overline{ e(x,\go,s)} \ d\gS$$
$$\quad =i {\rm lim}_{\ge\rightarrow 0}\sum_{x,y\  \in V_q}\int_{S^0 \times\gO}\overline{\hat{f}_{\rm sc}(\go,s)}\ W(x)
\overline{ G_0(\gl_{s}+i\ge)} (x,y) g(y) \overline{ e(x,\go,s)}
\  d\gS$$
$$\quad =i \sum_{y\in V_q}\int_{S^0\times\gO}\overline{\hat{f}_{\rm sc}(\go,s)}
[\sum_{x\in V_q} \overline{G_0(\gl_{s})} (x,y) W(x)\overline{
  e(x,\go,s)}]
 g(y) \ d\gS$$
$$\quad =i \sum_{y\in V_q}\int_{S^0 \times\gO}\overline{\hat{f}_{\rm sc}(\go,s)}
[\overline{e(y,\go,s)} - \overline{e_0(y,\go,s)}] g(y)\  d\gS$$
$$\quad =i (f,g) - i \int_{S^0 \times\gO}\overline{\hat{f}_{\rm sc}(\go,s)}
\hat{ g}(\go,s)\ d\gS . \quad\quad\quad\quad\quad\quad$$
In the second line above we used the Lippmann-Schwinger equation and
 at the last step we used the isometric property of the 
deformed Fourier Helgason transform (Theorem \ref{fhtil} ).
\end{demo}

We are ready to prove the following

\begin{theo}\label{tvers} 
For any $f$ and $g\in C_0(\T_q)$ 
\begin{equation*}(f, (S-I) g)=-2\pi i\int_{(S^0\times 
\gO)^2} T(\go,s;\go',s') \overline{\hat{ f}(\go,s)}
\delta (\gl_s-\gl_{s'})
 \hat{ g}(\go',s')\ d\gS d\gS'
\end{equation*}
where we set:  $d\gS d\gS': = d\gs_O(\go)
 d\mu (s)d\gs_O(\go') d\mu (s') $,  
(see Appendix A for the precise definition of the measure
 $\delta (\gl_s-\gl_{s'}) d\mu (s) d\mu (s')$).

This can be written symbolically by
\begin{equation}\label{tverse} 
S(\go,s;\go',s')= \gd(s-s')-2\pi i T(\go,s;\go',s')\gd(\gl_s-\gl_{s'})\ .
\end{equation}
\end{theo}
\begin{demo} From the definition of $S$ we get that
\begin{equation*}
(f, (S-I) g)= ((\Omega^- -\Omega^+)f,\Omega^+ g)=
{\rm lim}_{T\rightarrow \infty}\int_{T}^{-T}(e^{itA}(iW)e^{-itA_0}f,\Omega^+ 
g)dt
\end{equation*}
\begin{equation*}
 = (-i){\rm lim}_{\ge\rightarrow 0}
\int_{-\infty}^{+\infty}e^{-\ge| t|}(e^{itA}We^{-itA_0}f,\Omega^+ g)dt
 = (-i){\rm lim}_{\ge\rightarrow 0}\int_{-\infty}^{+\infty}
e^{-\ge| t|}{\cal L}(t)dt\ ,\quad\quad\quad\quad\quad
\end{equation*}
with\begin{equation}\label{calcul}
 {\cal L}(t)= \int_{S^0 \times \gO}\overline{{\cal FH}_{\rm sc}
(e^{itA}We^{-itA_0}f}) (\go',s'){\cal FH}_{\rm sc}
 (\Omega^+ g)(\go',s')\ d\gS'\ .\end{equation}
In the last step we used the isometric property of
 ${\cal FH}_{\rm sc} $ and the fact that $\Omega^+ g \in {\cal H}_{ac}$. 
Moreover we have 
$\Di{\cal FH}_{\rm sc}(\Omega^+ g)(\go',s')=\hat{ g}(\go',s')
$ (Lemma \ref{ffff})
and $\Di{\cal FH}_{\rm sc}(e^{itA}We^{-itA_0}f) (\go',s')=
e^{i\gl_{s'} t}{\cal FH}_{\rm sc}(We^{-itA_0}f) (\go',s')
$
$\Di\quad =e^{i\gl_{s'}t} \sum_{x\in V_q}(We^{-itA_0}f)(x)
\overline{ e(x,\go',s')}   $\\
$\Di\quad = \sum_{x,y\in V_q}\int_{S^0 \times \gO} e^{i(\gl_{s'}-\gl_s)
  t}
W(x,y) e_0(y,\go,s) \hat{ f}(\go,s)   \overline{ e(x,\go',s')} 
\ d\gS\ .$
Thus the expression in (\ref{calcul}) is
$${\cal L}(t)= \sum_{x,y\in V_q}\int_{(S^0\times 
\gO)^2}  e^{i(\gl_s-\gl_{s'}) t-\ge|
t|}\overline{V(x,y)}
\overline{ e_0(y,\go,s)}  \overline{\hat{ f}(\go,s)}   e(x,\go',s') 
\hat{ g}(\go',s') \ d\gS d\gS'  \,$$
and after doing the t-integration we get,
$$(f, (S-I) g)= (-i){\rm lim}_{\ge\rightarrow 0}\int_{(S^0\times 
\gO)^2} 
T(\go,s;\go',s')\frac{2\ge}{(\gl_s-\gl_{s'})^2+
 \ge^2} \overline{\hat{ f}(\go,s)}
 \hat{ g}(\go',s')   \ d\gS d\gS'  \ .$$
 We conclude by noticing as previously that the measures 
$$dl_\ge = \frac{ 2\ge  d\mu (s) d\mu (s') }{{(\gl_s-\gl_{s'})}^2 +\ge^2}$$
converge weakly to $2\pi\gd (\gl_s-\gl_{s'}) d\mu (s) d\mu (s')$
as $\ge \ra 0^+$. \end{demo}

A consequence of the relation between $T$ and $S$ is the unitarity
relation
 for $T$:
\begin{theo}\label{unitar}
Suppose $\ga \notin {\cal E}$. Then for any $s$ and $s'\in S^+$ with
 $\gl_s=\gl_{s'}=\ga$, and for any $(\go,\go')\in \gO\times\gO ,$
\begin{equation}\label{calc}\Im T(\go,s;\go',s') = \pi \int_{S^0 \times
\gO} \overline {T(\go'',s'';\go,s)}T(\go'',s'';\go',s')\gd
 (\gl_{s''}-\ga) d\gS'' 
\end{equation}
\end{theo}
\begin{demo}
By Theorem \ref{tvers} we have
$$\overline{\hat{(Sf)}} (\go,s) =\hat{f}(\go,s)-2\pi
i\int_{S^0 \times\gO}
 T(\go,s;\go',s')\hat{ f}(\go',s')\delta (\gl_s-\gl_{s'})
 d\gS'\ .
$$
The adjoint of the map $M :\hat{f}\ra \hat{(Sf)}$ is clearly given by
$$(M^*(g)) (\go,s)= g  (\go,s) +2\pi i \int_{S^0 \times\gO} \overline
 {T(\go',s';\go,s)}g(\go',s')\delta (\gl_s-\gl_{s'})
 d\gS'\ .
$$
The relation $M^*M=I$, which follows from $S^*S=I$, implies that
 (\ref{calc}) holds.
\end{demo}

\subsection{The S-matrix and the asymptotics of the deformed
 plane waves}
Next result explicits the link between the asymptotic behavior of the generalized eigenfunctions and
the coefficients 
of the scattering
 matrix.
\begin{theo}
There exist ``transmission coefficients'' $ \gt(s,\go,\go')$ so that
 the solution of the Lippmann-Schwinger equation (\ref{foitt}) writes 
$$ e(x;\go,s)=e_0(x;\go,s)+ \gt(s,\go,\go')q^{(-\ha +is)|x|}$$ for any
$x$
 close enough to $ \go' ~,$
 and these coefficients  are related to the scattering
 matrix by
the following
 formula 
\begin{equation}\label{trco}
S(\go',-s;\go,s)= -\frac{2 i \pi}{C(s)}  \gt(s,\go,\go') 
\end{equation}
with$\quad \Di C(s) =\frac{1}{q^{\ha -is}-q^{-\ha +is}}\ .$
\end{theo}
\begin{demo}
From the study of the Lippmann-Schwinger equation (\ref{foitt}),
 we write  the decomposition
\[ e(x;\go,s)=e_0(x;\go,s)+e_{\rm scatt}(x;\go,s) ~,\]
where 
\[ e_{\rm scatt}(x;\go,s)=\sum _{y\in K }
G_0(\gl_s,x,y)g(y;\go,s)~\]
 where   $g(y;\go,s)=\sum _{z\in K} W(y,z)e(z,\go,s)$.\\
Let us look at the asymptotic behaviour of $e_{\rm scatt}(x;\go,s)$
as $x\ra \go'$.
We have seen (Theorem \ref{greenarbre}) that the Green's function  
$G_0(\gl_s;x,y)$  satisfies equation (\ref{mspi})
\[ G_0(\gl_s;x,y)=C(s)q^{(-\ha +is)d(x,y) }; \] 
%and $|d(x,y)-|x||\leq |y|$; 
then (\ref{futtt}) and (\ref{rad}) imply 
that, if $x \ra \go'$,
\[ e_{\rm scatt}(x;\go,s)=\gt(s,\go,\go')q^{(-\ha +is)|x|}~,\]
with
% ce ne serait pas mieux de citer le lemma 2.1 ?
\[  \gt (s,\go,\go')=C(s)\sum_{y\in K} g(y;\go,s)q^{(\ha
  -is)b_{\go'}(y)}
 = C(s)\sum_{(y,z)\in K\times K}e(z,\go,s)  \overline{W(z,y)} e_0(y,\go',s) ~.\]
Noticing that $ e_0(y,\go',s)= \overline{e_0(y,\go',-s)}$ we get that 
\[ \gt (s,\go,\go')= C(s)T(\go',-s;\go,s) ~.\]
and from (\ref{tverse}) we derive formula (\ref{trco}).\end{demo}
\begin{rem}
For any $y \in K$ we have  $b_\go (y)= b_{\go'}(y)$ 
if $\go $ and $\go'$ belong to the same end of  $\T_q \setminus K$.
This implies that the function $ \go'\rightarrow \gt(s,\go,\go')$
 is in fact constant
in each end of $\T_q \setminus K$, so that the transmission coefficient
$\gt(s,\go,\go')$ can be written as a function $\gt(s,\go,l)$.
Moreover the reduced Lippmann-Schwinger equation 
depends only on the restriction
of $e_0$ to $K$, this implies that the function 
$\go \rightarrow \gt(s,\go,l)$ is also constant  in each end  
 of $\T_q \setminus K$.
Finally, we get an $L\times L$ matrix depending on $s$, denoted
by $${ \tilde S}(s) =\left( S(l',-s,l,s)\right)_{l,l'}=
-\frac{2 i \pi}{C(s)}\left(\gt(s,l,l')\right)_{l,l'}$$. 
\end{rem}
\subsection{Computation of the transmission coefficients in terms of the Dirichlet-to Neumann operator}

In this section, we compute the transmission coefficients
 following the method of \cite{S}. Let us recall that 
\[ K=\{ x\in V_q~|~\exists y \in V_q {\rm ~ with ~}W_{x,y}\ne 0 \}~.\] 

 We recall that  $|x|$ denotes the combinatorial distance of the vertex
$x$ to the root $O$ of $\T_q$. Let us set $B_{n-1}=\{ x\in V_\gG ~|~ |x|
 \leq n-1 \}$, where $n$ is chosen so that 
$n-2$ is the supremum of $|x|$ for  $x$ in $K$.
 We  denote by $T_l$ the ends of $\T_q \setminus B_{n-1}$ ($1\leq l\leq L$),
 by  $x_l$  the root  of $T_l$ and  by $\gO_l$
 the boundary of $T_l$, which consists in the set of all
geodesic rays starting  from  $x_l$ and staying into $T_l $.
 The set of the roots
  $\{ x_l~|~ l=1,\cdots, L\}$ is the circle of radius $n$
 and $L= (q+1)q^{n-1}$.
\begin{figure}[hbtp]
\leavevmode \center
\input{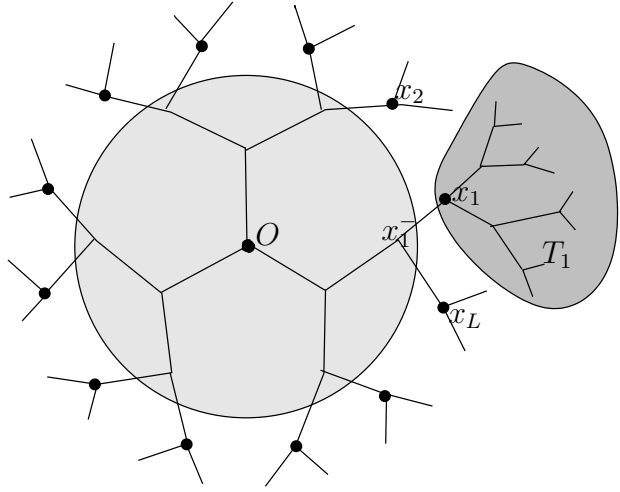}
\caption{{\it \small The tree $\T_2$, the  ball $B_{n-1}$ and the end $T_1$ for $n=3$.}}
 \label{fig:tree}
\end{figure}
From now on, we consider a fixed $l$ ($1\leq l\leq L$),
 a fixed geodesic ray $\go$ in
$\gO_{l}$,
 and the associated "incoming plane wave"
\begin{equation*} \label{plw} 
 \forall x \in V_q, \quad e_0(x,\go,s) = q^{(1/2-is)b_{\go}(x)},
 \end{equation*}
 where
 $s\in S^0$.
We recall that such a plane wave is a generalized eigenfunction for the adjacency
operator
 $A_0$ on $\T_q $ in the sense that it satisfies 
\begin{equation*} \label{scat} 
  (\lambda_s-A_0)e_0(x,\go,s)=0\ ~ (\lambda_s=2\sqrt q  \cos (s\log q))~,
 \end{equation*} 
but is not in $l^2$.
We are looking for solutions 
\begin{equation*} \label{splw} 
  \quad  e(.;\go,s)= e_{0}(.;\go,s)+ e_{scat}(.;\go,s),\quad x \in V_q
 \end{equation*}
of the equation 
\begin{equation}\label{helm} 
  \quad (\lambda_s-A)e(.;\go,s) = 0 ,
 \end{equation} where  the scattered wave $e_{scat}(.;\go,s)$
satisfies:
\begin{equation}\label{holm} e_{scat}(x,\go,s)= \gt(s,\go,l')\Phi_s
  (x)
 \quad {\rm if} \quad x \in V(T_{l'})~, \end{equation}
where \begin{equation*} \label{radial} 
   \Phi_s(x) = q^{(-1/2+is)|x|}\ 
 \end{equation*}
(the so-called radiation
condition)
 and the coefficients $\gt(s,\go,l')$ are the  transmission
 coefficients. 
 These radial waves are generalized eigenfunctions of $A_0$
 in the sense 
defined previously.
We want to get an explicit expression of the  transmission vector
\begin{equation} \label{vector} \overrightarrow {\gt(s,\go)}:=
 ({\bf\gt}(s,\go,1),\cdots, \gt(s,\go,l'),\cdots,
   {\bf\gt}(s,\go,L)). \end{equation}
As we shall
see, the  transmission vector  does not depend on the choice of the
 geodesic ray $\go$, 
it is uniquely determined by the choice of $l$; we define 
\begin{equation} \label{trv} \overrightarrow{\gt(s,l)}:= 
\overrightarrow{\bf\gt(s,\go)}, 
 \quad \forall \go \in \gO_l ~.\end{equation}
 We thus recover the result of the previous section, with
the following relation
 $$\forall l,l' \quad \gt(s,l,l')=-\frac{C(s)}{2i\pi}
 S(l',-s,l,s)\ .$$
We begin with noticing that $ b_{\go}(x_{l'})$ does not depend of
 $\go \in \gO_l$ for any $l' \in \{1,\cdots,L\}$. 
We set
 \begin{equation*} \label{vectora} \overrightarrow{{ A}_l}:=
(\ga^{-b_{\go}(x_{1})},\cdots, 
 \ga^{-b_{\go}(x_{L})}) =  (\ga^{-b_{l}(x_{1})},\cdots, 
 \ga^{-b_{l}(x_{L})}) \quad (\ga =q^{-1/2+is}), \end{equation*}
and denote by $\overrightarrow{E_{l}}$  the vector in $\R^L$ 
having all null coordinates excepted the $l$-th coordinate,
 which is equal to $1$.\\ 
We will prove the following 
\begin{theo}
Consider  the integer $n$ so that $B_{n-2}$ is the smallest ball
 containing the finite graph $K$.

Set $\gG= B_n$,  $\partial\gG = \{x_{l'},1\leq l'\leq L \}$,
denote by  $\widehat{ A_n}$ the restriction of ${ A}$  to $B_{n}$ in
 the sense that $\widehat{ A_n} =(A_{x,y})_{(x,y)\in B_n}$,
 define $I_n$ in the same way,
 set $B= \widehat{ A_n}- \lambda_s I_n$ and denote
 by ${\cal DN}_s$ the corresponding Dirichlet-to Neumann operator
 (see Definition \ref{dton}, Appendix B).

Then ${\cal DN}_s$ and the transmission vector
 $\overrightarrow{\gt(s,l)}$ defined by (\ref{holm}),(\ref{vector})
 and (\ref{trv}) 
exist for any $$s \notin {\cal E}_0= \{s\in S^0\ ;\ 
 \lambda_s  \in \sigma ( \widehat{A}_{n-1})\}$$ and
$$\left( \gt(s,l,l')\right)=-\ga^{-2n}\left[
 \frac{1}{C(s)}\left({\cal DN}_s + q^{1/2+is}I \right)^{-1} +
 {\cal A}\right]\ ,  $$
with $\widehat{ A}_{n-1}= (A_{x,y})_{(x,y)\in B_{n-1}} \    $,
  $\ {\cal A}=({\cal A}_{l,l'})=(\ga^{d(x_l,x_{l'})}), 
\quad \ga =q^{-1/2+is}\ .$
\end{theo}
\begin{demo}
Let us recall that we have fixed $l$ ($1\leq l\leq L$) and
 a geodesic ray $\go$ in
$\gO_{l}$. From now on we  write $e(x)$ instead
 of $e(x,\omega,s)$  for any $x \in V_q$ for
simplicity.\\
Equation (\ref{helm}) splits into 3 expressions, depending
 on where $x$ is taken.
\begin{itemize}
\item If $x \notin B_n$ 
 the equation is already verified, since
 $A$ coincide with $A_0$ on each end $ T_{l'}$. 
\item if $x \in B_{n-1}$, the equation becomes
 the Dirichlet problem  
$\Di {A} e(x) =\gl_s e(x) , \quad x \in B_{n-1}$,
$\Di e_{|\partial\gG} = f\ ,$
and is uniquely solvable for any $s$ outside ${\cal E}_0$,
  the prescribed values of the function $f$ at the boundary being given 
by the $L$-vector 
\begin{equation*} \vec {e_l}=:(e(x_1)... e(x_L)) \end{equation*}
to be determined in the sequel.
\item if $x \in \partial\gG = \{x_{l'},1\leq l'\leq L \}= B_n \setminus B_{n-1}$, 
  then $x$ is one of the roots  $x_{l'}$ , and  
 equation (\ref{helm}) writes:
\begin{equation} \label{Front}
\forall l' \in \{ 1,\cdots ,L\},
\quad  e(x_{l'}^-) + \sum_{x\sim x_{l'},x\in T_{l'}} e(x) = \lambda_s e(x_{l'})
\end{equation}
where  $x_{l'}^-$ is the unique interior neighbor of $x_{l'}$, (see figure \ref{fig:tree}),
and where we used that the potential $W$ vanishes outside $B_{n-2}$.
 \end{itemize}
According to definition \ref{dton} (Appendix B), the 
 Dirichlet-to Neumann operator  ${\cal DN}_s$ corresponding to $B= \widehat{ A_n}- \lambda_s I_n$ writes 
\begin{equation*}
 {\cal DN}_s(\vec {e_l})(x_{l'}) =  e(x_{l'}^-) -\lambda_s e(x_{l'})
 \quad \forall l'
 \in \{ 1,\cdots,  L\} ,\end{equation*}
%where we denote by ${\bf e}^{L}$ the $L$-vector$ (e(x_1)... e(x_L))\ .$ 

Therefore, if $s\notin {\cal E}_0$,  (\ref{Front})
 can be rewritten as follows
\begin{equation} \label{frint}
\forall l' \in \{ 1,\cdots ,L\}, 
\quad  {\cal DN}_s(\vec {e_l})(x_{l'}) + \sum_{x\sim x_{l'},x\in T_{l'}} e(x) = 0
\end{equation}

Now it remains to compute $\sum_{x\sim x_{l'},x\in T_{l'}} e(x)$.
 We have, for $ x\in T_{l'}$, 
$ e(x)= q^{(1/2-is)b_{\go}(x)} +
  \gt(s,\go,l')  \Phi_s (x) $.
According to the expression of the radial function $ \Phi_s$,
we get, for any $s\notin {\cal E}_0$ and
  $l' \in \{ 1,\cdots ,L\}$, that
\begin{eqnarray*}
 \Phi_s (x_{l'})&=& q^{(-1/2+is)|x_{l'}|}=q^{n(-1/2+is)} \\
 \Phi_s (x)&=&q^{(-1/2+is)(n+1)}\quad\forall x\in T_{l'} \quad   x\sim x_{l'} .
\end{eqnarray*}
Let us write the set $ N_l= \{x\in T_{l},
\quad x\sim x_l\} $ as $N_l= \{y_l\} \cup \tilde{ N_l}$,
where $y_l$ belongs to the infinite path $\go$ whereas the $q-1$
 vertices of $\tilde { N_l}$ do not. 
Then, using  the properties of the Busemann function we have
\begin{eqnarray*}
 b_{\go}(x_{l})&=& b_{l}(x_{l})=|x_{l}|=n \\
 b_{\go}(y_l)&=& n+1 \\  
 b_{\go}(x)&=& b_{l}(x_{l})-1 =n-1, \quad\forall  x\in \tilde{N_l}\\
 b_{\go}(x)&=&  b_{l}(x_{l'})-1\quad \forall  x\in N_{l'}, 
 \quad \quad l' \neq l ~.
\end{eqnarray*}
So we get for any   $l' \in \{ 1,\cdots ,L\}$ and 
 after setting $\ga =q^{-1/2+is}$, 
\begin{eqnarray*} \label{fronta}
e(x_{l'})&=&  \ga^{-b_{l}(x_{l'})}+ \gt(s,\go,l') \ga^{n}\\
e(x)&=&   \ga^{-b_{l}(x_{l'})+\ge}+ \gt(s,\go,l') \ga^{n+1}\end{eqnarray*}
with \begin{eqnarray*}\ge &= &1 \quad \forall  x\in \tilde{N_l} \\
\ge &=& 1, \quad \forall \ x\in  N_{l'}\quad l' \neq l\\
\ge& =&-1 \quad {\rm if}\quad x=y_{l} \ .\end{eqnarray*}
Hence we have
$$\sum_{x\sim x_{l'},x\in T_{l'}} e(x)= q\ga^{-b_{l}(x_{l'})+1}+
 q\gt(s,\go,l') \ga^{n+1}\quad {\rm if}\quad l' \neq l
$$
$$\sum_{x\sim x_{l},x\in T_{l}} e(x)= (q-1)\ga^{-b_{l}(x_{l})+1}+
 \ga^{-b_{l}(x_{l})-1}+q\gt(s,\go,l') \ga^{n+1}\ .
$$
These equations can be summarised,
 for any  $l' \in \{ 1,\cdots ,L\}$, as
$$\sum_{x\sim x_{l'},x\in T_{l'}} e(x)=
 \ga^{-b_{l}(x_{l'})+1}[q+ \delta_{ll'}(\ga^{-2}-1)]+q\gt(s,\go,l') \ga^{n+1}
$$
so that   equation (\ref{frint}) gives, for any  $l' \in \{ 1,\cdots ,L\}$ 
\begin{equation} \label{fronte}
{\cal DN}_s(\vec {e_l})(x_{l'}) + \ga^{-b_{l}(x_{l'})+1}[q+
 \delta_{ll'}(\ga^{-2}-1)]+q\gt(s,\go,l') \ga^{n+1}  = 0
\ .
\end{equation}
Let us set ${\cal DN}_s(\vec {e_l})=
 ({\cal DN}_s(\vec {e_l})(x_1),...{\cal DN}_s(\vec {e_l})(x_L))$,\\
and  write $\vec {e_l}= (e(x_1),...e(x_L))= {\vec A}_l + 
 \ga^{n} \overrightarrow{\gt(s,\go)}$\ (recall that  \\
${\vec A}_l$ and $\overrightarrow{\gt(s,\go)}$
 are $L-$vectors having respectively $
 \ga^{-b_{l}(x_{l'})}$ and $\gt(s,\go,l')$  as their $l'-$coordinate).\\
 
Substituting in (\ref{fronte}) and denoting by
 $\vec{ E_{l}}$  the vector in $\R^L$
having all null coordinates except $x_{l}=1$,  we get
$$ {\cal DN}_s[{\vec A}_l +  \ga^{n}
 \overrightarrow{\gt(s,\go)}]+q\ga {\vec A}_l +
 \ga^{-n}(\ga^{-1}-\ga)\vec{ E_{l}} +q\overrightarrow{\gt(s,\go)} \ga^{n+1}=
0\,$$
which yields
$$\ga^n\left({\cal DN}_s  +
 q\ga I\right)\overrightarrow{\gt(s,\go)} =
 \ga^{-n}(\ga-\ga^{-1})\vec{ E_{l}} -\left(q\ga I +{\cal DN}_s 
\right){\vec A}_l\ .$$
Using the expression of $\ga $ and $C(s)$, we have then
$$\ga^n\left( {\cal DN}_s +
  q^{1/2+is}I\right)\overrightarrow{\gt(s,\go)} =
 -\frac{\ga^{-n}}{C(s)}\vec{ E_{l}} -\left(q^{1/2+is}I + 
 {\cal DN}_s\right){\vec A}_{l}\ .$$
Since the matrix $ {\cal DN}_s$ is  real symmetric,
  $  {\cal DN}_s + q^{1/2+is}I$ is an invertible matrix for any $s \in S^0$ 
 so that $\lambda_s \notin \sigma ( \widehat{A}_{n-1})$,
 and
$$\overrightarrow{\gt(s,\go)}=
 -\frac{\ga^{-2n}}{C(s)}\left({\cal DN}_s +
 q^{1/2+is}I \right)^{-1} \vec{ E_{l}}-\ga^{-n}{\vec A}_{l}\ .  $$
We conclude the proof by noticing that,
 for any  $l' \in \{ 1,\cdots ,L\}$,   $b_{l}(x_{l'})=n- d(x_l, x_{l'})$.
\end{demo}

{\it Acknowledgment: we thank  the referee for his careful reading
of our manuscript and for suggesting many improvements to our initial
text.}

\section*{Appendix A: delta measures}

The goal of this Appendix is to define in a precise way
the meaning of the measures 
$d\mu =\delta (S=0) d\nu $
where $d\nu =a(x) dx $ is absolutely continuous w.r. to the
Lebesgue measure in $\R^d$  and $S$ is a $C^1$ real valued function
so that $dS $ does not vanish on the hyper-surface $S=0$.
The measure $d\mu =\delta (S=0) d\nu $ is supported by the 
 hyper-surface $\Sigma:=\{ S=0 \}$.

We can assume that $\R^d$
and  the hyper-surface $S=0$ are oriented, so that we can play with
differential forms instead of measures.

The proof of the following Lemma is left to the reader:
\begin{lemm} There exists a differential form
$\beta $ defined in some neighborhood of $\Sigma $
so that $adx_1 \wedge \cdots \wedge dx_d = dS \wedge  \beta $.
Moreover the restriction of $\beta $ to $\Sigma $ is uniquely
defined.
\end{lemm}
\begin{defi} If $\nu $ is the measure on $\Sigma $
  associated to the restriction
of $\beta $ to $\Sigma $. we can  view $\nu $ as a measure on $\R^d$
denoted
$d\nu= \delta (S=0)d\mu $.
\end{defi}

We can view $d\nu $ as weak limits:
if $f :\R \ra \R^+ $ is a positive $L^1$ function of integral $1$
and $f_\ge (t)=\ge^{-1} f(\ge^{-1}t )$
the measure
$\delta (S=0) d\mu $ is the weak limit as $\ge \ra 0 $
of the measures $d\nu _\ge = f_\ge (t) d\mu $ 
(For the proof, take local coordinates so that $S=x_1$).

Usual choices are $f_1$ the characteristic function of the interval
$[-\ha,\ha]$ and 
$f_2(t)=\frac{1}{\pi} \frac{1}{1+t^2}$.

\section*{Appendix B: the Dirichlet-to-Neumann operator
 $\cal{DN}$ on a finite graph}

Let  $\gG=(V,E)$ be a connected finite graph
 and let $\partial\gG $ be a subset of $V$ called 
 the "boundary of $\gG$".
 Let $B=(b_{i,j}): \ \R^V\rightarrow\R^V$ be a symmetric matrix
 associated
 to $\gG$, namely
$$b_{i,j}=0 \quad {\rm if~} i\neq j\ {\rm ~and~}
 \quad\{i,j\} \notin E.$$
After setting $V_0= V\setminus \pa \gG$, we define
 $B_0 : \ \R^{V_0}\rightarrow\R^{V_0}$ as the restriction of $B$ to
the functions which vanish on $\partial\gG $ .

We have the following 
\begin{lemm}\label{dirch} 
Assume that $B_0$ is invertible.
Then, for any given  $f \in C (\partial\gG )$, there exists a unique solution
  $F \in  C (\gG )$
 of the  Dirichlet problem 
\[ (D_f)~:~ \Di F_{|\partial\gG} = f~{\rm and~} 
\Di BF(l)=0 {~\rm if ~} l\in V_0
\ .\]
\end{lemm}

The {\it Dirichlet-to-Neumann}
 operator ${\cal DN}$ associated to $B$ is the linear operator
 from $C (\partial\gG )$ to $C (\partial\gG )$ 
 defined as follows:
\begin{defi}\label{dton}
Assume that $B_0$ is invertible.
Let $f \in C (\partial\gG )$,
 and $F$ be the unique solution of the Dirichlet problem $(D_f)$.
Then, the Dirichlet-to-Neumann operator form
${\cal DN}:\R^{\pa \Gamma} \ra \R^{\pa \Gamma}  $
is defined as follows:  if $l\in \partial\gG$, 
$${\cal DN}(f)(l) = \sum_{i=1}^{m} b_{l,i} F(i)(=BF(l))
 ~ .$$
\end{defi}


\begin{thebibliography}{666}


\bibitem[1]{Bre} J. Breuer.
 \newblock {\it Singular continuous spectrum
 for the Laplacian on certain sparse trees.}
 \newblock Commun. Math. Phys. {\bf 269 (3)}:851--857 (2007).

\bibitem[2]{Cartier}
P. Cartier,
      {\it G\'eom\'etrie et analyse sur les arbres,}
  S\'eminaire Bourbaki, 24\`eme ann\'ee (1971/1972), Exp.
              No. 407,
  Lecture Notes in Math. Springer, {\bf 317}:123--140 (1973).
   
    

\bibitem[3]{CdV1} Y. Colin de Verdi\`ere.
\newblock {\it Spectre de graphes.}
\newblock Cours sp\'ecialis\'es {\bf  4},
 Soci\'et\'e math\'ematique de France (1998).


\bibitem[4]{CdV2} Y. Colin de Verdi\`ere.
{\it Distribution de points sur une sph\`ere.}
 S\'eminaire N. Bourbaki,  expos\'e  {\bf 703}:83--93 (1988-89).

\bibitem[5]{CdV3} Y. Colin de Verdi\`ere.
{\it Mathematical models for passive imaging 
I: general background.}
\newblock  ArXiv 0610043.

\bibitem[6]{CdV4} Y. Colin de Verdi\`ere.
{\it Semiclassical analysis and passive imaging.}
  \newblock Nonlinearity{\bf 22}:45--75 (2009).

\bibitem[7]{CdV5} Y. Colin de Verdi\`ere.
{\it A Semi-classical calculus of correlations.}
  \newblock  Thematic issue ``Imaging and Monitoring with Seismic Noise''
 of the series ``Comptes Rendus G\'eosciences'',
 from the Acad\'emie des sciences {\bf 343}:496--501 (2011).

\bibitem[8]{CMS} M. Cowling, S. Meda \& A. Setti.
\newblock {\it An overview of harmonic analysis on the
 group of isometries of a 
regular tree. }
\newblock Exposition.Math., {\bf 16(5)}:385--423
(1998).

\bibitem[9]{CS} M. Cowling \& A. Setti.
\newblock {\it The range of the Helgason-Fourier transformation on 
regular trees. }
\newblock Bull.Austral.Math.Soc., {\bf 59}:237--246
(1998).

\bibitem[10]{FN} A. Fig\`a-Talamanca \& C. Nebbia,
\newblock {\it Harmonic Analysis and representation theory for
groups acting on regular trees. }
\newblock London Math. Soc. Lecture Notes Series, {\bf 162}
Cambridge Univ. Press, 1991.

\bibitem[11]{Ike} T. Ikebe.
{\it Eigenfunction expansion associated with the Schr\"odinger
  operators an their applications to scattering theory.}
Arch. Rational Mech. Anal., {\bf 5}:1--34 (1960).


\bibitem[12]{RS} M. Reed \&  B. Simon.
\newblock{\it Methods of Modern mathematical Physics}
 III-{\it Scattering theory,} (1980),
\newblock New York, Academic Press.


\bibitem[13]{S} U. Smilansky
\newblock{\it Exterior-Interior Duality for Discrete Graphs}
J. Phys. A: Math. Theor., {\bf 42}:035101 (2009). 



\end{thebibliography}
\end{document}